\documentclass[journal]{IEEEtran}
\usepackage{silence}
\usepackage[utf8]{inputenc}
\usepackage[T1]{fontenc}
\usepackage{cite}
\usepackage{amsmath,amssymb,amsfonts,mathtools}
\usepackage{graphicx}
\usepackage{textcomp}
\usepackage{xcolor}
\usepackage{enumerate}
\usepackage{adjustbox}
\usepackage{setspace}
\usepackage{needspace}
\usepackage{algorithm}
\usepackage{tikz}
\usetikzlibrary{patterns,calc}
\usepackage{pgfplots}
\pgfplotsset{compat=newest}

\newcommand{\shift}{\operatorname{sft}}
\newcommand{\shiftt}{\operatorname{SFT}}

\newcommand{\Bo}{B^{\mathrm{o}}}
\newcommand{\Bi}{B^{\mathrm{i}}}
\newcommand{\ko}{\ell^{\mathrm{o}}}
\newcommand{\ki}{\ell^{\mathrm{i}}}
\newcommand{\Eo}{\mathcal{E}^{\mathrm{o}}}
\newcommand{\Ei}{\mathcal{E}^{\mathrm{i}}}

\newcommand{\qed}{\hfill$\square$}
\newcommand{\klo}{\underline{\kappa}}
\newcommand{\khi}{\overline{\kappa}}

\newtheorem{assumption}{Assumption}

\newtheorem{lemma}{Lemma}
\newtheorem{theorem}{Theorem}
\newtheorem{remark}{Remark}
\newtheorem{proposition}{Proposition}
\newtheorem{corollary}{Corollary}

\newenvironment{pf}{\par\noindent\textit{Proof:} }{\par}

\def\BibTeX{{\rm B\kern-.05em{\sc i\kern-.025em b}\kern-.08em
    T\kern-.1667em\lower.7ex\hbox{E}\kern-.125emX}}

\begin{document}
\title{Shifted Interaction Functions: A Unified Framework for Imperfect Actuation and Sensing in Consensus Networks}

\author{\centerline{Yang Xu, Enric Morella Violeta, and H\'ector Garc\'ia de Marina}%
\thanks{This work was supported by the ERC Starting Grant iSwarm (101076091) and the RYC2020-030090-I grant from the Spanish Ministry of Science.}%
\thanks{The authors are with the Department of Computer Engineering, Automation and Robotics, and the Institute of Mathematics of Granada, University of Granada, Granada, Spain.}%
\thanks{Corresponding author: H\'ector Garc\'ia de Marina (e-mail: hgdemarina@ugr.es).}%
}

\maketitle

\begin{abstract}
Practical networked robotic systems rarely implement ideal consensus interactions exactly, since local actuation constraints, sensing biases, and heterogeneous interaction maps may shift the input-output relations used by distributed controllers. Rather than focusing solely on whether consensus occurs, such imperfections can fundamentally alter the structure of the asymptotic collective behavior, typically yielding global or clustered velocity alignment, while exact static agreement arises only in special cases. This paper develops a shifted-interaction framework for analyzing collective alignment under imperfect actuation and sensing. Shifted functions provide a compact representation of local actuator and sensor maps, allowing actuation constraints and sensing biases to be represented by horizontally and vertically shifted monotone interactions. The framework distinguishes nodal shifted actuation from edge-wise shifted sensing. For shifted actuation, a Lyapunov analysis establishes global velocity alignment under compatible actuation ranges, including bounded, non-invertible, and deadzone-type maps. For shifted sensing, a maximum-principle argument and a paired-sensitivity condition establish convergence to velocity clusters over edges with infinite accumulated sensitivity; uniform sensing sensitivity yields exponential global velocity alignment. In the linear case, closed-form expressions characterize asymptotic behavior and identify actuation as a special case of sensing. Simulations illustrate static, globally aligned, and clustered velocity-aligned behaviors.
\end{abstract}


\section{Introduction}
\label{sec:introduction}

\IEEEPARstart{D}{istributed} consensus algorithms are a fundamental tool for cooperative control and coordination in multi-agent systems. Classical protocols are usually built on idealized interactions in which agents apply unconstrained actuation, exchange symmetric relative-state information, and use coupling functions that vanish at the origin \cite{bullo2018lectures}. In robotic implementations, however, actuators may saturate or work only in one direction, sensors may contain calibration biases, and agents or edges may have different local limits and gains. Such heterogeneous imperfections arise in fixed-wing unmanned aerial vehicles without reverse thrust, where non-negative shifted saturation has recently been validated in outdoor formation-flight experiments \cite{xu2025distributed}, in satellites with on-off thrusters \cite{zhang2015satellite}, and in robotic networks affected by biased or asymmetric relative measurements. Remark~\ref{rem_higher} explains how far our first-order coordination model covers these systems.

Several strands of the literature address related issues, but mostly as separate problems. Consensus under input saturation \cite{yang2014global} and under saturated interconnection states \cite{lim2015consensus} has been studied in both classical and recent settings. Distributed and networked extensions further consider nonidentical bounds, asymmetric limits, and connectivity-preserving interactions \cite{zou2023consensus, wang2024consensus, zhu2025distributed}. Shifted actuation or sensing ranges can exclude the origin and induce nonzero collective velocities, unlike the origin-admissible saturation models in these studies. State- or output-constrained consensus and safety-aware formulations also show that bounded interactions can change collective behavior \cite{fu2023safe, lin2023constrained, wang2023connectivity}. Sensor bias in consensus is usually handled from a compensation viewpoint, where biased relative measurements are estimated or rejected rather than used to characterize the motion they induce \cite{sinha2022consensus}. Nonlinear edge-function networks have been analyzed through monotonicity and passivity tools \cite{chen2019convergence}; their reciprocal couplings yield static, possibly clustered, steady states, whereas our distinct, offset oriented maps need not admit equilibria and may produce drifting velocity clusters. Closest in spirit, mismatched relative measurements in rigid formations generate distorted shapes and residual motions \cite{demarina2015controlling, sun2014formation} and have been exploited for maneuvering \cite{demarina2016distributed, de2021maneuvering}. These studies draw on the gradient structure of distance-based formations \cite{demarina2015controlling, sun2014formation, demarina2016distributed} or linear displacement-based consensus \cite{de2021maneuvering}, the latter fitting linear shifted sensing. Unlike these formation-specific mechanisms, we compare shifts acting after nodal aggregation with those acting before aggregation along oriented edges. Our shifted-actuation result and sensing maximum principle include non-invertible maps, whereas our velocity-clustering result assumes $C^1$ sensing and comparable paired sensitivities, which deadzones or saturations need not satisfy. Together, these results motivate a consensus-level description of saturation, bias, unilateral actuation, and shifted sensing within a common interaction model.

This paper develops such a description by modeling actuation and relative sensing through linear or nonlinear horizontally and vertically shifted monotone interaction functions. A conventional consensus coupling is an origin-vanishing map of a state or relative-state error. We generalize this map by allowing horizontal and vertical shifts, so that actuator constraints, one-sided actuation, sensor biases, and asymmetric edge couplings can be represented by the same mathematical object. The framework is applied to two complementary channels. In nodal shifted actuation, the aggregate consensus input is transformed before driving each agent. In edge-wise shifted sensing, each measured relative state is transformed before aggregation. Although the notation starts from an underlying undirected interaction graph, each physical edge is treated through two oriented channels, allowing heterogeneous and effectively directed local processing. Actuator limits lead to range-compatibility checks, sensing nonlinearities lead to responsiveness conditions on active edges, and linear shifted maps provide closed-form predictions of drift and limiting offsets. In the linear case, nodal actuation is representable by oriented-edge sensing, but arbitrary heterogeneous sensing gains need not admit a nodal actuation representation. For nonlinear interactions, the location of a shift may change velocity alignment, the steady-state offset, and whether the limiting behavior is static or drifting.

Motivated by these channel-dependent effects, the contributions of this paper are summarized as follows:
\begin{itemize}
\item[1)] A shifted-interaction formulation places actuator saturation, unilateral actuation, sensor bias, dead zones, and asymmetric couplings within a unified consensus model that separates nodal shifted actuation from edge-wise shifted sensing.
\item[2)] For shifted actuation, a range-compatibility condition guarantees global velocity alignment for invertible, bounded, and non-invertible maps, and characterizes the asymptotic motion; static consensus is a nongeneric special case.
\item[3)] For shifted sensing, a maximum-principle analysis is organized through an active-graph viewpoint. Under a paired-sensitivity condition, individual velocities converge and edges with infinite accumulated sensitivity yield cluster-level alignment; uniform sensing sensitivity yields exponential global alignment.
\item[4)] For linear shifts, closed-form expressions give the drift velocity and the limiting configuration, and linear shifted actuation is shown to be a special case of linear shifted sensing, a correspondence lost for nonlinear interactions.
\end{itemize}

Section~\ref{sec:notation} introduces notation, Section~\ref{sec:protocols} reviews the consensus problem, and Section~\ref{sec:modeling} formulates shifted actuation and sensing. Sections~\ref{sec:actuation}--\ref{sec:linear} analyze shifted actuation, shifted sensing, and the linear case, Section~\ref{sec:simulation} presents simulations, and Section~\ref{sec:conclusion} concludes the paper.

\section{Notation}
\label{sec:notation}
The $n$-dimensional identity matrix is denoted by $I_n$, and $\mathbf{1}_n$ represents the all-ones vector. The operator $\operatorname{diag}(\cdot)$ creates a diagonal matrix with its arguments on the diagonal. For a matrix $A$, the notation $A^\dagger$ stands for its Moore–Penrose pseudoinverse, and $\ker(A)$ and $\operatorname{im}(A)$ denote its kernel and image. The operator $\operatorname{col}(\cdot)$ stacks its arguments into a column vector, and $\otimes$ denotes the Kronecker product. For a scalar function $f$ and a value $y$, $f^{-1}(y):=\{s\in\mathbb{R} : f(s)=y\}$ denotes the preimage set, which is a singleton when $f$ is invertible. We write $\varphi(t)=o(\chi(t))$ if $\varphi(t)/\chi(t)\to0$ as $t\to\infty$.

\begin{figure}[!htbp]
\centering
\makebox[\columnwidth][c]{\includegraphics[page=1,width=0.35\columnwidth]{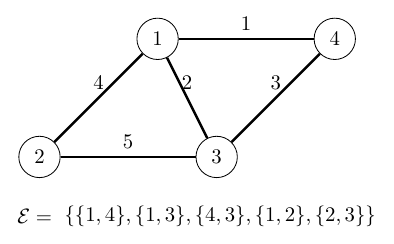}}\par\vspace{0.8mm}
\makebox[\columnwidth][c]{\includegraphics[page=2,width=0.35\columnwidth]{img_graph/graph.pdf}\hspace{0.025\columnwidth}\includegraphics[page=3,width=0.35\columnwidth]{img_graph/graph.pdf}}

\caption{Example of a bidirectional graph to model asymmetric interactions. For each bidirectional interaction in $\mathcal{E}$, the \emph{positive} and \emph{negative} orientations are chosen arbitrarily.}
\label{fig:graph}
\end{figure}

In this paper, we consider $N$ agents with $M$ bidirectional interactions. These interactions are encoded by a graph $\mathcal{G} = (\mathcal{V}, \mathcal{E})$, where each agent is associated with a node in the set $\mathcal{V} = \{1, \ldots, N\}$, and each bidirectional interaction is represented by an edge in the set $\mathcal{E} \subseteq \{\{i,j\} : i,j\in\mathcal{V},\ i\neq j\}$. The edge set $\mathcal{E}$ has cardinality $M$ and consists of unordered pairs $\{i,j\}$ when agents $i$ and $j$ interact.%
We index the edges by $\ell$ and use $\ell \in \mathcal{E}$ and $\{i,j\} \in \mathcal{E}$ interchangeably; agents $i$ and $j$ are then \emph{neighbors}. Although information is exchanged in both directions, each agent may process it differently due to heterogeneous sensing, actuation, or local interaction maps. It is therefore convenient to consider both orientations $(i,j)$ and $(j,i)$ of each edge separately\footnote{Note that the one-direction and the bidirectional interactions are represented by $(i,j)$ and $\{i,j\}$, respectively.}. We arbitrarily assign one orientation as the \emph{positive} one, denoted by $\ki$, and the other as the \emph{negative} one, denoted by $\ko$; the sets $\Ei$ and $\Eo$ collect the positively and negatively oriented edges, and we sometimes omit the superscript of $\ell$. For each node $i$, $\mathcal{N}_i$ is the set of its neighbors, $\mathcal{E}_i$ is the set of oriented edges with initial node $i$, i.e., the interactions from which agent $i$ receives information, and $\mathcal{E}_i^\mathrm{i} = \mathcal{E}_i \cap \Ei $, $\mathcal{E}_i^\mathrm{o} = \mathcal{E}_i \cap \Eo$; see Figure~\ref{fig:graph}. 

The incidence matrix $B \in \mathbb{R}^{N \times M}$ is defined by%
$[B]_{i\ell} = +1 \,(-1)$ if node $i$ is the initial (terminal) node of edge $\ell\in \Ei$, and $0$ otherwise; the \emph{Laplacian} is $L = BB^\top$ and the \emph{edge Laplacian} is $L_e = B^\top B$. We also define the out- and in-incidence matrices $\Bo,\Bi\in\mathbb{R}^{N\times M}$: $[\Bo]_{i\ell}=+1$ if $i$ is the initial node of edge $\ell\in\Ei$, and $[\Bi]_{i\ell}=-1$ if $i$ is the terminal node, and 0 otherwise for both matrices. Note $B = \Bo + \Bi$. The superscripts of $\Bo,\Bi$ refer to the role (out/in) of a node in a positively oriented edge, whereas those of $\Ei,\Eo$ label the orientation; thus the measurement along $\ki$, taken by the initial node, enters through $\Bo$, and that along $\ko$ through $\Bi$. We will always consider the following assumption.
\begin{assumption}
\label{ass_graph}
$\mathcal{G}$ is connected and the interactions by two neighboring agents are bidirectional.
\end{assumption}

\begin{remark}
We stress that Assumption~\ref{ass_graph} goes beyond a simple undirected graph with symmetric interactions between neighbors. Indeed, it accounts for two distinct oriented sensing channels per edge, rendering the graph directed in practice.
\end{remark}

\section{Consensus protocols and paper's objective}
\label{sec:protocols}

Consensus protocols focus on the following limits of the agents' states $x_i\in\mathbb{R}, i\in\mathcal{V}$
\begin{equation}
\lim_{t\to\infty} (x_i(t)-x_j(t))  \quad \forall \{i,j\}\in\mathcal{E}, \label{eq_lim_xixj}
\end{equation}
particularly the system is said to have achieved \textit{consensus} when the previous limits are all equal to zero. For a connected, balanced, and  undirected graph, this is equivalent to the convergence onto the consensus manifold $\mathcal{M} := \ker(L) = \ker(B^\top)$, spanned by $\mathbf{1}_N$, as described in \cite{bullo2018lectures}. Let us consider single-integrator agents 
\begin{equation}
\label{eq: si}
    \dot x_i(t) = u_i(t), \quad i\in\mathcal{V},
\end{equation}
where $u_i \in\mathbb{R}$ is the control input. The \emph{standard} consensus algorithm \cite{bullo2018lectures} is
\begin{equation}\label{equ_std_consensus}
    u_i(t) = - \sum_{j\in\mathcal{N}_i} (x_i(t) - x_j(t) ), \quad i \in \mathcal{V},
\end{equation}
which achieves (average) consensus if $\mathcal{G}$ is undirected and connected, i.e., $\lim_{t\to\infty} x_i(t) = \frac{1}{|\mathcal{V}|}\sum_{j\in\mathcal{V}} x_j(0)$; importantly, such a consensus implies \emph{staticity}, i.e., $\lim_{t\rightarrow\infty} \dot{x}_i(t) =0, \forall i\in\mathcal{V}$.

In multi-robot systems, where $x_i$ may be a position or an orientation, configurations other than \emph{static consensus} are also of interest, such as \emph{drifting consensus}, where consensus is achieved but a residual velocity remains, $\lim_{t\rightarrow\infty} \dot{x}_i(t) \neq 0$. This motivates studying \emph{velocity alignment} through the limits
\begin{equation}
\lim_{t\to\infty} (\dot{x}_i(t)-\dot{x}_j(t)) \quad \forall \{i,j\}\in\mathcal{E},
\label{eq_lim_dotxixj}
\end{equation}
where agents' velocities are \emph{aligned} if all \eqref{eq_lim_dotxixj} equal zero.

This paper studies \eqref{eq_lim_xixj} and \eqref{eq_lim_dotxixj} when actuation and sensing use heterogeneous shifts of origin-crossing functions, e.g., biased measurements and positive-only actuation.

\enlargethispage{10pt}
\begin{remark}\label{rem_higher}
\textit{Higher-Order Alignment -} 
For higher-order systems than (\ref{eq: si}), velocity alignment (\ref{eq_lim_dotxixj}) can extend to acceleration or higher-order. Studying whether these alignments are nonzero is critical in robotics, as \emph{nonstaticity} may damage actuators or drive the system out of functioning configurations. The single integrator \eqref{eq: si} models the coordination layer of the agents. For instance, in the fixed-wing formation flights of \cite{xu2025distributed}, the consensus input acts on the amplitude of an oscillatory guidance maneuver; this amplitude is always non-negative and drives a positive progression speed along the path, so that the map from the consensus input to the progression speed is a non-negative shifted saturation. For second-order agents, a related mechanism may appear one derivative order higher; Section~\ref{sec:simulation} illustrates this only for linear shifts. Discontinuous on--off thrust does not satisfy Assumption~\ref{ass_shift}. When such agents have a velocity-tracking loop obeying the stated hypotheses, the results apply to that coordination layer. For states $x_i\in\mathbb{R}^d$ with axis-wise shifts and a common orientation of the agents' coordinate frames, coordinate-separable nodal actuation and oriented sensing maps each yield $d$ scalar problems over the same graph, so the corresponding results apply per coordinate. For coupled gradient-type actuation $F_i=\nabla\phi_i$, with $\phi_i$ convex and $F_i$ locally Lipschitz, $\eta:=(L\otimes I_d)x$ and $V=\sum_i\phi_i(-\eta_i)$ give $\dot V=-\mathbf F^\top(L\otimes I_d)\mathbf F\le0$, $\mathbf F:=\operatorname{col}(F_i(-\eta_i))$; the proof of Theorem~\ref{thm_inp_constrained2} carries over if $\{\eta\in\operatorname{im}(L\otimes I_d):\mathbf F\in\ker(L\otimes I_d)\}$ is nonempty and bounded. General coupled sensing maps are not covered.
\end{remark}

\Needspace{7\baselineskip}
\section{Shifted actuation and sensing modeling}\label{sec:modeling}

\subsection{Practical challenges in consensus implementation}
\label{subsec:practical}
Physically implementing a consensus algorithm requires each agent to \textit{sense} the relative states of its neighbors, \textit{process} them into a control input, and \textit{actuate} this input.

\begin{figure}[!htbp]
    \centering
    \includegraphics[width=0.85\linewidth]{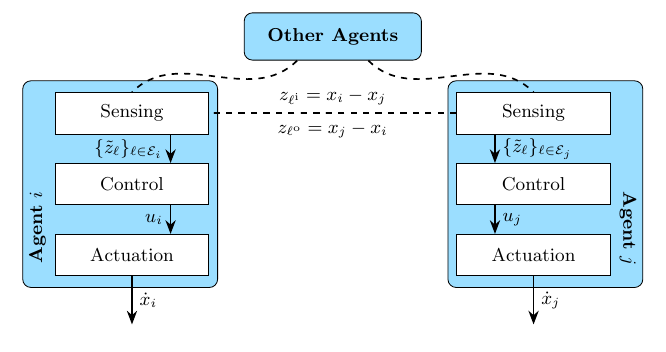}
    \caption{Schematic representation of the implementation of a consensus algorithms, such as (\ref{equ_std_consensus}), involving the $\ell$-th edge $\{i,j\}\in\mathcal{E}$. Dashed lines indicate bidirectional measurements.}
    \label{fig_modules}
\end{figure}

Consider, as in Figure~\ref{fig_modules}, neighbors $i$ and $j$ connected by the edge $\ell\in\mathcal{E}$, with $\ki=(i,j)$ and $\ko=(j,i)$. Agents $i$ and $j$ only sense $z_{\ki} := x_{i} - x_{j}$ and $z_{\ko} := x_{j} - x_{i}$, respectively. Note that we can say that $z_{\ki}$ ($z_{\ko}$) is the $\ki$-th ($\ko$-th) relative state concerning the set $\mathcal{E}_i^\mathrm{i}$ ($\mathcal{E}_j^\mathrm{o}$, since $\ko=(j,i)$ has agent $j$ as its initial node).
The sensing produces the measurements $\tilde{z}_{\ki} = \psi_{\ki}(z_{\ki})$ and $\tilde{z}_{\ko} = \psi_{\ko}(z_{\ko})$, for some functions $\psi_{\ki}, \psi_{\ko}: \mathbb{R} \to \mathbb{R}$ modeling imperfections. Following \eqref{equ_std_consensus}, each agent aggregates the measurements over $\mathcal{E}_i = \Ei_i \cup \Eo_i$, i.e.,
\begin{equation}\label{equ_actual_control}
u_i = - \sum_{\ki\in\mathcal{E}_i^\mathrm{i}} \tilde{z}_{\ki} - \sum_{\ko\in\mathcal{E}_i^\mathrm{o}} \tilde{z}_{\ko}.
\end{equation}

The input \eqref{equ_actual_control} is then executed by the actuation module. With perfect measurements, $\tilde{z}_{\ell} = z_{\ell}$, and faithful actuation, $\dot x_i=u_i$, the \emph{standard} protocol \eqref{equ_std_consensus} is recovered. Physical multi-robot systems, however, face misaligned or miscalibrated components; high-precision hardware mitigates these effects but limits scalability. Conversely, understanding these imperfections allows exploiting shifted sensing or actuation intentionally.

\subsection{Formalization of consensus with shifted functions}
\label{subsec:system}

We model deterministic, usually unknown, and heterogeneous imperfections as
\begin{equation}
\label{eq: func}
\dot{x}_i = F_i(u_i), \;
\tilde{z}_{\ell} =\psi_{\ell}(z_{\ell}),%
\end{equation}
for some functions $F_i,\psi_{\ell}:\mathbb{R}\to\mathbb{R}$ and for all nodes $i\in\mathcal{V}$ and oriented edges $\ell\in \Ei \cup\Eo$.%
These functions are called \textit{shift functions} and are detailed in Subsection~\ref{subsec:shift}.

\begin{figure}[!htbp]
    \centering
    \includegraphics[width=0.75\linewidth]{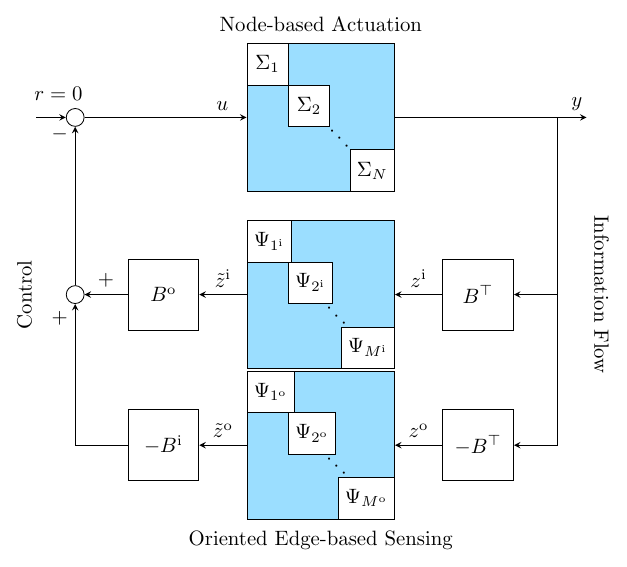}
    \caption{Control-theoretical description of a distributed consensus protocol in order to analyze and accommodate shift functions in actuation and relative sensing.}
    \label{fig_controlplant}
\end{figure}

As in \cite{chen2019convergence}, we describe the closed loop compactly in Figure~\ref{fig_controlplant}: first-order systems $\Sigma_i$ at the nodes, with null references $r_i=0$, and measurement systems $\Psi_\ell$ on the oriented edges $\ell\in \Ei \cup \Eo$, whose direction coincides with (is opposite to) that of $B$ for $\ell\in\Ei$ ($\ell\in\Eo$). Stacking the signals, $x = [x_1,\cdots, x_N]^\top$, $y = x$, $z^\mathrm{i} = B^\top y$, $z^\mathrm{o} = - B^\top y$, $\tilde{z}^\mathrm{i} = [\tilde{z}_{1^\mathrm{i}}, \cdots, \tilde{z}_{M^\mathrm{i}}]^\top$ and $\tilde{z}^\mathrm{o} = [\tilde{z}_{1^\mathrm{o}}, \cdots, \tilde{z}_{M^\mathrm{o}}]^\top$, and $u = [u_1, \dots, u_N]^\top$, the input in Figure~\ref{fig_controlplant} is
\begin{equation}
\label{equ_compact_control}
u = - \Bo \tilde{z}^\mathrm{i} + \Bi \tilde{z}^\mathrm{o}.
\end{equation}

Finally, the systems in Figure~\ref{fig_controlplant} are modelled by
\begin{equation}
    \label{equ_general_system}
    \Sigma_i :  \begin{cases} 
        \dot{x}_i = F_i(u_i), \\
        y_i = x_i,
    \end{cases} \quad 
    \Psi_\ell :
        \tilde{z}_{\ell} = \psi_{\ell}(z_{\ell}), \\
\end{equation}
for some shift functions $F_i, \psi_{\ell}:\mathbb{R}\rightarrow\mathbb{R}$, for all nodes $i\in\mathcal{V}$ and oriented edges $\ell\in\Ei \cup\Eo$. Indeed, the \emph{standard} protocol \eqref{equ_std_consensus} with \eqref{eq: si} is recovered when all shift functions are the identity.

\subsection{Shift Functions}\label{subsec:shift}

\begin{figure}[!htbp]
\centering
\includegraphics[width=\columnwidth]{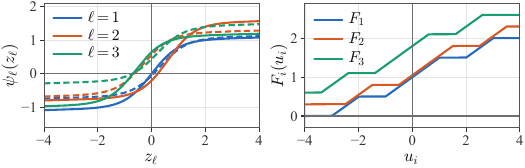}
\caption{The left subfigure shows six shifted sensing functions associated with three bidirectional interactions, one color per interaction, with solid and dashed lines for its two orientations. The nominal sensing model $\psi_\ell$ in \eqref{equ_general_system} is a $\operatorname{tanh}$-type sigmoid, but each $\psi_\ell$, for $\ell\in\Ei \cup \Eo$, differs by scaling and by horizontal and vertical shifts representing sensing imperfections. The right subfigure shows three shifted actuation models $F_i$ in \eqref{equ_general_system}. These actuation functions are piecewise linear and non-negative by design, with flat intervals representing jammed or saturated actuation.}
\label{fig: sftact}
\end{figure}

The term \emph{shift function} refers to shifting, vertically and horizontally, the non-decreasing origin-vanishing functions used in consensus, as illustrated in Figure~\ref{fig: sftact}. For example, a miscalibrated sensor might shift input and output as $\shift(s) = a f(s - h) + b$, where $a>0$ is a scaling factor, $h\in\mathbb{R}$ a horizontal shift, $b\in\mathbb{R}$ a vertical shift, and $f:\mathbb{R}\to\mathbb{R}$ is an increasing function vanishing at the origin. The lowercase symbol $\shift$ always denotes a scalar local shift function, such as $\shift_i$ at a node or $\shift_{ij}$ on an oriented measurement; in particular, $\shift_i\equiv F_i$ and $\shift_\ell\equiv\psi_\ell$ in \eqref{equ_general_system}, and the figures keep the labels $F_i$ for the actuation maps. The uppercase symbol $\shiftt$ denotes the stacked vector map, e.g., $\shiftt(u)=\operatorname{col}(\shift_1(u_1),\ldots,\shift_N(u_N))$.

\begin{assumption}
\label{ass_shift}
\textit{Monotone Shift Functions -} 
All shift functions $\shift(\cdot):\mathbb{R}\to\mathbb{R}$ are locally Lipschitz continuous and monotonically non-decreasing.
\end{assumption}

Assumption~\ref{ass_shift} covers linear, sigmoid-like, and piecewise monotone maps with dead zones or flat saturations, and the value at zero need not be zero. Local Lipschitz regularity supplies classical well-posedness and the almost-everywhere differentiability needed in subsequent arguments; the comparison proof of Theorem~\ref{thm_active_graph}(i) itself uses only continuity and monotonicity. The role of shift parameters $a$, $h$, and $b$ is made explicit in Subsection~\ref{subsec:role} and Section~\ref{sec:linear}.

\section{Consensus under shifted actuation}
\label{sec:actuation}

\subsection{Requirements for the analysis of shifted actuation}

Let $R_i:=\shift_i(\mathbb{R})$, an interval by Assumption~\ref{ass_shift}.

\begin{assumption}\label{ass_range}
\textit{Compatible Actuation Ranges -} 
The ranges of all shift functions are compatible, i.e., the set $\mathcal{R} := \bigcap_{i\in\mathcal{V}} R_i$ is nonempty and has non-zero length.
\end{assumption}

Assumption~\ref{ass_range} is a sufficient range-compatibility condition for velocity alignment with finite compatible inputs; asymptotic alignment may also occur at range-closure endpoints. The non-zero length condition is discussed after Theorem~\ref{thm_inp_constrained2}.

\subsection{Shifted Actuation System Description}
For shifted actuation, \eqref{equ_general_system} becomes
\begin{equation}\label{equ_input_system}
    \Sigma_i :  \begin{cases} 
        \dot{x}_i = \shift_i(u_i), \\
        y_i = x_i,
    \end{cases} \quad 
    \Psi_\ell :
    \,
        \tilde{z}_{\ell} = z_{\ell},
\end{equation}
which with \eqref{equ_compact_control} gives $u = - (\Bo + \Bi )B^\top y = - Lx$ and
\begin{equation}
\label{equ_shift_input}
\dot{x} = \shiftt(-Lx),
\end{equation}
We define $\eta := Lx\in\operatorname{im}(L) = \mathrm{span}(\mathbf{1}_N)^\perp$, a subspace isomorphic to $\mathbb{R}^{N-1}$.

\subsection{Shifted Actuation Analysis}

We recall LaSalle's Invariance Principle~\cite{khalil2002nonlinear} as a lemma and three auxiliary lemmas.
\begin{lemma} \label{LaSalleThm2}
\textit{LaSalle Invariance Principle (\cite{khalil2002nonlinear}, Theorem 4.4) -}
     Let $\Omega \subset \mathcal{D}$ be a compact set that is positively invariant with respect to an autonomous first-order system $\dot x = f(x)$ with $f:\mathcal{D}\to\mathbb{R}^n$ locally Lipschitz. Let $V: \mathcal{D} \to \mathbb{R}$ be a continuously differentiable function such that $\dot{V}(x) \leq 0$ in $\Omega$. Let $E$ be the set of all points in $\Omega$ where $\dot{V}(x) = 0$. Let $\mathcal{I}$ be the largest invariant set in $E$. Then every solution starting in $\Omega$ approaches $\mathcal{I}$ as $t \to \infty$.
\end{lemma}

\begin{lemma}\label{lemma_onesol_nonginv}
\textit{Unique Compatible Velocity -} 
    Under Assumptions~\ref{ass_shift} and \ref{ass_range} there exists exactly one $y^*\in\mathcal{R}$ such that $\sum_{i\in\mathcal{V}} \eta_i = 0, \eta_i \in -\shift_i^{-1}(y^*)$ for some $\eta_1,\dots,\eta_N\in\mathbb{R}$, where $\shift_i^{-1}(y^*)$ is the preimage set.
\end{lemma}

\begin{lemma}\label{lemma_convexity}
\textit{Separable Convexity Criterion -} 
    Let $f:\mathbb{R}^k\to\mathbb{R}$ be a continuously differentiable function for some $k$. Suppose that each component of the gradient $\frac{\partial f}{\partial x_i}$ is a non-decreasing function of $x_i$ alone, then $f$ is convex.
\end{lemma}
\begin{lemma}\label{lemma_radially}
\textit{Radial Unboundedness Test -} 
    Let $f:\mathbb{R}^k\to\mathbb{R}$ be a continuously differentiable, convex function for some $k$. Suppose that its set of critical points $\mathcal{X}^\star:=\{\xi\in\mathbb{R}^k : \nabla f(\xi)=\mathbf{0}_k\}$ is nonempty and bounded. Then each $\xi^\star\in\mathcal{X}^\star$ is a global minimizer of $f$, and $f$ is radially unbounded.
\end{lemma}
For the respective proofs, see Appendices \ref{proof: lemma_onesol_nonginv}, \ref{proof: lyaconvex} and \ref{proof: lyaunbounded}.

\begin{theorem}\label{thm_inp_constrained2}
\textit{Shifted Actuation Alignment -} 
Consider the multi-agent system \eqref{equ_input_system} whose graph $\mathcal{G}$ satisfies Assumption~\ref{ass_graph}, and the $\shift$ functions in the actuation satisfy Assumptions~\ref{ass_shift} and \ref{ass_range}. The system achieves velocity alignment, i.e., $\lim_{t\to\infty}\dot x(t) \in \operatorname{span}\{\mathbf{1}_N\}$. More precisely, $\lim_{t\to\infty}\dot x(t) = y^\ast\mathbf{1}_N$, where $y^\ast$ is the unique value given by Lemma~\ref{lemma_onesol_nonginv}. Moreover, $\eta(t)=Lx(t)\to\bar\eta\in\operatorname{im}(L)$ with $\shiftt(-\bar\eta)=y^\ast\mathbf1_N$; this limit may depend on $x(0)$.
\end{theorem}
\begin{pf}
We show that $\eta=Lx$ converges to one equilibrium $\bar\eta$ satisfying $\shiftt(-\bar\eta)=y^\ast\mathbf1_N$, implying velocity alignment. Consider the evolution of $\eta$, which is governed by the autonomous first-order system
\begin{equation}\label{equ_difeta}
\dot{\eta}= L\dot{x}= L\shiftt(-\eta),
\end{equation}
By construction, $\eta$ clearly stays in $\operatorname{im}(L)$. The right-hand side of \eqref{equ_difeta} is locally Lipschitz by Assumption~\ref{ass_shift}, so its solutions exist and are unique.
    
(Lyapunov function and monotonicity) Consider the following candidate Lyapunov function $V:\operatorname{im}(L)\to\mathbb{R}$
\begin{equation}\label{equ_Vdef}
V(\eta) = \sum_{i\in\mathcal{V}} \int_{0}^{-\eta_i} \shift_i(s) \,\mathrm{d}s,
\end{equation}
which is continuously differentiable because each integrand is continuous. Note that $V$ is non-increasing along trajectories 
\begin{equation}
\begin{aligned}\label{equ_derivV1}
    \dot{V}(\eta(t)) &=  - \sum_{i\in\mathcal{V}} \shift_i(-\eta_i(t))\dot{\eta}_i(t) \\&= - \shiftt(-\eta(t))^{\top} L \shiftt(-\eta(t)) \leq 0,
\end{aligned}
\end{equation}
for all $t\geq 0$, because $L$ is positive semi-definite. 

(Boundedness and non-emptiness of the set $E$) We now study the set $E = \{ \eta \in \operatorname{im}(L) \, \vert \, \dot{V}(\eta) = 0 \}$. The condition $\eta\in\operatorname{im}(L)$ implies $\mathbf{1}_N^\top\eta = 0$, or, equivalently $\sum_{i\in\mathcal{V}}\eta_i=0$. On the other hand $\dot{V}=0$ can only be satisfied if $\shiftt(-\eta)\in\ker(L) = \mathrm{span}(\mathbf{1}_N)$, i.e., $\shift_i(-\eta_i) = y^*, \ \forall i\in\mathcal{V}$ for some $y^*\in\mathcal{R}$. Lemma~\ref{lemma_onesol_nonginv} guarantees that there exist exactly one $y^*\in\mathcal{R}$ such that the previous two conditions hold simultaneously. 

The corresponding values of $\eta$ at which $\shiftt(-\eta) = y^\ast\mathbf{1}_N$ may be non-unique, because the shift functions are not necessarily invertible; however, they must remain bounded. Indeed, suppose $E$ was unbounded, then there would exist a sequence of points in $E$ with at least two unbounded components, one unbounded above (say, index $i_0$) and the other unbounded below (say, index $i_1$), because $\sum_{i\in\mathcal{V}} \eta_i = 0$. However, due to Assumption~\ref{ass_shift} (increasingness of the shift functions), this would imply $\shift_{i_0}(-\eta_{i_0}) \ge y^\ast$, for all $\eta_{i_0} \in \mathbb{R}$, and $\shift_{i_1}(-\eta_{i_1}) \le y^\ast$, for all $\eta_{i_1} \in \mathbb{R}$, which contradicts Assumption~\ref{ass_range}, as it would force $\mathcal{R} \subset R_{i_0}\cap R_{i_1} = \{ y^\ast\}$ to be empty or have zero length. Hence, $E$ must be bounded.

(Compactness of the set, $\Omega$)  Furthermore, we show that $V$ is radially unbounded. First, note that $V$ is the restriction of the function $\widetilde{V}:\mathbb{R}^N\to\mathbb{R}$ to $\operatorname{im}(L)$, where $\widetilde{V}$ is calculated as $V$ in \eqref{equ_Vdef} but defined on a larger domain. 
Note that $\partial\widetilde{V}/\partial w_i(w) = -\shift_i(-w_i)$, i.e., $\nabla \widetilde{V} (w) = -\shiftt(-w)$ for all $w\in\mathbb{R}^N$. Each component is a non-decreasing function of $w_i$ alone, so by Lemma~\ref{lemma_convexity},
$\widetilde{V}$ is convex, and so is its restriction $V$. Additionally, the gradient of $V$ within $\operatorname{im}(L)$ is $\nabla V (\eta) = -\Pi\, \shiftt (-\eta)$, where $\Pi := I_N - \frac{1}{N}\mathbf{1}_N\mathbf{1}_N^\top$ is the orthogonal projection of $\mathbb{R}^N$ onto $\operatorname{im}(L)$. Note that $\nabla V(\eta)=0$ if and only if $\shiftt(-\eta)\in\ker(\Pi)=\mathrm{span}(\mathbf{1}_N)$; therefore, points where the gradient vanishes correspond to solutions to $\sum_{i\in\mathcal{V}} \eta_i = 0, \eta_i \in -\shift_i^{-1}(y^*)$, i.e., points in $E$. According to Lemma~\ref{lemma_onesol_nonginv} the set $E$ is non-empty and by a previous argument bounded, thus by Lemma~\ref{lemma_radially}, applied on $\operatorname{im}(L)\cong\mathbb{R}^{N-1}$, $V$ is radially unbounded. Finally, we can define the compact and positively invariant set $\Omega = \{ \eta \in \operatorname{im}(L) \, \vert \, V(\eta) \le V(\eta(0)) \}$; indeed, it is positively invariant by \eqref{equ_derivV1} and compact because of the radial unboundedness of $V$.

(Largest invariant set) Every $e\in E$ is an equilibrium of \eqref{equ_difeta}. LaSalle's principle gives $\operatorname{dist}(\eta(t),E)\to0$. To show pointwise convergence, for $e\in E$ set $D_e:=(\eta-e)^\top L^\dagger(\eta-e)$. Since $\eta-e\in\operatorname{im}(L)$ and $L^\dagger L=\Pi$,
\[
\dot D_e=2\sum_i(\eta_i-e_i)\big(\shift_i(-\eta_i)-\shift_i(-e_i)\big)\le0
\]
by monotonicity. Choose an accumulation point $\bar\eta\in E$, which exists by compactness of $\Omega$ and LaSalle's principle. Then $D_{\bar\eta}(t)$ is non-increasing and tends to zero along a subsequence, so $D_{\bar\eta}(t)\to0$. Positive definiteness of $L^\dagger$ on $\operatorname{im}(L)$ yields $\eta(t)\to\bar\eta$. The limit may depend on the initial condition when $E$ is not a singleton; nevertheless, $\dot x(t)=\shiftt(-\eta(t))\to y^\ast\mathbf1_N$. Thus velocity alignment holds. \qed
\end{pf}

The non-zero length condition in Assumption~\ref{ass_range} prevents upper and lower saturation levels of different agents from coinciding, in which case $V$ would not be radially unbounded. It also excludes constant actuation, typical of leaders. If exactly one agent $i_0$ has constant actuation $\dot{x}_{i_0} = v_{\mathrm a}$, with $v_{\mathrm a}$ in the range of every other agent but not one of its saturation levels, the same arguments show that all agents attain the velocity $v_{\mathrm a}$; the case of a fixed agent, $v_{\mathrm a}=0$, is detailed in Corollary~\ref{coro_anchor}.

Global invertibility makes the compatible input vector uniquely determined by the shift maps.

\begin{assumption}\label{ass_ginv}
\textit{Globally Invertible Shifts -} 
    Shift functions are globally invertible.
\end{assumption}

Under Theorem~\ref{thm_inp_constrained2}, let $\bar\eta:=\lim_{t\to\infty}\eta(t)\in E$, which may depend on $x(0)$ if the maps are non-invertible. Then $\Pi x(t)=L^\dagger\eta(t)\to L^\dagger\bar\eta$. Moreover, without invertibility or a rate assumption,
\[
x(t)=L^\dagger\bar\eta+y^\ast t\mathbf 1_N+\alpha_\infty\mathbf 1_N+o(1).
\]
Indeed, set $q:=x-L^\dagger\bar\eta-y^\ast t\mathbf1_N$. At a maximum of $q_i$, $(Lq)_i\ge0$ and $\dot q_i=\shift_i(-\bar\eta_i-(Lq)_i)-y^\ast\le0$; the reverse inequality holds at a minimum. Thus $\max_i q_i$ is non-increasing and $\min_i q_i$ is non-decreasing. As $\Pi q=L^\dagger(\eta-\bar\eta)\to0$, both approach the same constant $\alpha_\infty$. Under Assumption~\ref{ass_ginv}, $E$ is a singleton and $\bar\eta_i=-\shift_i^{-1}(y^\ast)$, making the relative offset explicit.

The following corollary characterizes when (static) consensus is achieved.
\begin{corollary}\label{coro_realconsensus}
\textit{Static Consensus Recovery -} 
 In Theorem~\ref{thm_inp_constrained2} with the additional Assumption~\ref{ass_ginv}, consensus among the states $x_i(t), \forall i \in\mathcal{V}$ is achieved
if and only if all shift functions intersect at zero input, i.e., $\shift_i(0) = \shift_j(0), \forall i,j\in\mathcal{V}$. Moreover, the consensus is static if and only if all shift functions pass through the origin, $\shift_i(0) = 0, \forall i\in\mathcal{V}$.
\end{corollary}

\begin{pf}
(Sufficiency) Suppose $\shift_i(0) = \shift_j(0)$ for all $i,j\in\mathcal{V}$. Then $y^* = \shift_i(0)$ satisfies the conditions of Lemma~\ref{lemma_onesol_nonginv} with $\eta_i=0$, and by the lemma it is the only such value. Moreover, by global invertibility $\bar{\eta} = \mathbf{0}_N$, implying $\Pi x(t)\to L^\dagger\bar\eta=\mathbf{0}_N$, and consensus is achieved eventually. If additionally, all shift functions pass through the origin, i.e., $\shift_i(0)=0, \forall i\in\mathcal{V}$, then $y^\ast = 0$ and knowing that $\lim_{t\to\infty}\dot{x}(t) = y^\ast \mathbf{1}_N$ (Theorem \ref{thm_inp_constrained2}), then the staticity $\lim_{t\to\infty} \dot{x}(t) = 0$ follows.

(Necessity) Suppose consensus is asymptotically achieved, i.e., $\Pi x(t)\to\mathbf{0}_N$; then $\eta(t)=L\Pi x(t)\to\mathbf{0}_N$, so $\bar{\eta}=\mathbf{0}_N$. The steady-state velocity of each agent is $\dot{x}_i = \shift_i(-\bar{\eta}_i)=\shift_i(0), \forall i\in\mathcal{V}$. By Theorem~\ref{thm_inp_constrained2} velocity alignment is achieved, therefore $\shift_i(0) = \shift_j(0), \forall i,j\in\mathcal{V}$. If staticity is additionally achieved then it must be $\dot x_i = \shift_i(0) = 0$ for all $i\in\mathcal{V}$. \qed
\end{pf}

Corollary~\ref{coro_realconsensus} shows that imperfections, even in a single agent, typically yield nonzero velocities and non-consensus states, so static consensus is nongeneric. Conversely, shifted actuation can be designed to induce a predictable collective motion, since known actuation maps determine $\lim_{t\to\infty}\dot x(t)$ and $\lim_{t\to\infty} \Pi x(t)$.

An anchored network has a fixed agent $i_0$ with $\shift_{i_0}\equiv0$. For $i\neq i_0$, boundedness of $\shift_i^{-1}(0)$ excludes an unbounded zero-valued saturation plateau. Under invertibility, set $\bar\eta_i=-\shift_i^{-1}(0)$ for $i\neq i_0$ and $\bar\eta_{i_0}=-\sum_{i\neq i_0}\bar\eta_i$, so $\bar\eta\in\operatorname{im}(L)$.

\begin{corollary}\label{coro_anchor}
\textit{Anchored Network -} Consider system~\eqref{equ_input_system} under Assumptions~\ref{ass_graph} and~\ref{ass_shift}, with $\shift_{i_0}\equiv0$. Suppose $0\in R_i$ and $\shift_i^{-1}(0)$ is bounded for all $i\neq i_0$. Then $\dot x(t)\to\mathbf 0_N$ and $x(t)$ converges to a finite limit, possibly dependent on the initial condition. If all remaining shift functions are globally invertible, then
\[
 x(t)\to x^\ast:=L^\dagger\bar\eta+\big(x_{i_0}(0)-(L^\dagger\bar\eta)_{i_0}\big)\mathbf 1_N.
\]
Moreover, $x^\ast=x_{i_0}(0)\mathbf 1_N$ if and only if $\shift_i(0)=0$ for all $i\neq i_0$.
\end{corollary}

\begin{pf}
Here Assumption~\ref{ass_range} fails, since $\mathcal R=\{0\}$, but the proof of Theorem~\ref{thm_inp_constrained2} applies with $y^\ast=0$. The function $V$ in \eqref{equ_Vdef} is unchanged, except that its $i_0$-th term vanishes, so \eqref{equ_derivV1} holds and $V$ is convex. Since the $i_0$-th entry of $\shiftt$ is zero, its critical points in $\operatorname{im}(L)$ form the set $E=\{\eta\in\operatorname{im}(L):\shift_i(-\eta_i)=0,\ i\neq i_0\}$. Each $\eta_i$, $i\neq i_0$, lies in the nonempty bounded set $-\shift_i^{-1}(0)$, and $\eta_{i_0}=-\sum_{i\neq i_0}\eta_i$, so $E$ is nonempty and bounded. Lemmas~\ref{lemma_radially} and~\ref{LaSalleThm2} then give $\dot x(t)\to\mathbf 0_N$. The monotone-distance argument of Theorem~\ref{thm_inp_constrained2} applies to the compact equilibrium set $E$ and yields $\eta(t)\to e\in E$. As the anchor is fixed, $x(t)\to L^\dagger e+[x_{i_0}(0)-(L^\dagger e)_{i_0}]\mathbf1_N$. Under invertibility, $E=\{\bar\eta\}$ and $e=\bar\eta$, giving the stated $x^\ast$. Finally, $L^\dagger$ is injective on $\operatorname{im}(L)$, which meets $\operatorname{span}(\mathbf 1_N)$ only at the origin, so $x^\ast\in\operatorname{span}(\mathbf 1_N)$ if and only if $\bar\eta=\mathbf 0_N$, i.e., $\shift_i(0)=0$ for all $i\neq i_0$; then $x^\ast=x_{i_0}(0)\mathbf 1_N$. \qed
\end{pf}

Corollary~\ref{coro_anchor} shows that fixing a single agent turns drift into a static distortion. The limiting offsets are determined by an element of $E$, and become explicitly computable as $L^\dagger\bar\eta$ under invertibility.

\subsection{Role of the shift parameters}\label{subsec:role}
For the parametric family $\shift_i(s) = a_i f_i(s-h_i)+b_i$, with $a_i>0$ and $f_i$ continuous, strictly increasing, and $f_i(0)=0$, the role of each parameter becomes explicit. The range is $R_i = b_i + a_i f_i(\mathbb{R})$: vertical shifts translate the range and the scaling stretches it, whereas horizontal shifts leave it unchanged. Hence horizontal shifts never affect the feasibility condition in Assumption~\ref{ass_range}, while vertical shifts may violate it when $f_i$ is bounded. The common velocity $y^\ast$ is the unique root of
\begin{equation}\label{eq:role}
\textstyle\sum_{i\in\mathcal{V}} \big[h_i + f_i^{-1}\big((y-b_i)/a_i\big)\big]=0,
\end{equation}
and the limiting aggregate inputs are $\bar\eta_i = -h_i - f_i^{-1}((y^\ast-b_i)/a_i)$, which set the state offsets $L^\dagger\bar\eta$. For instance, with purely horizontal shifts ($b_i=0$, $a_i=a$, $f_i=f$), \eqref{eq:role} gives $y^\ast = a f(-\bar h)$ with $\bar h := \frac1N\sum_i h_i$, and $\bar\eta_i = \bar h - h_i$: the average horizontal shift sets the drift velocity through $f$, while deviations from the average set the non-consensus offsets independently of $f$. With purely vertical shifts ($h_i=0$, $a_i=a$, $f_i=f$), $y^\ast$ solves $\sum_i f^{-1}((y-b_i)/a)=0$, which differs in general from the average $\frac1N\sum_i b_i$ obtained for linear $f$. This asymmetry disappears in the linear case of Section~\ref{sec:linear}.

\begin{remark}
\textit{Infeasible Range Case -} 
    If the closures of the ranges $R_i$ have no common point, velocity alignment is impossible, since a common limiting velocity would belong to the closure of every $R_i$. The Lyapunov decrease $\dot V\le0$ persists, but the coercivity argument used in Theorem~\ref{thm_inp_constrained2} no longer applies. The degenerate case in which $\mathcal R$ reduces to a point because one agent is fixed is covered by Corollary~\ref{coro_anchor}.
\end{remark}

\section{Consensus under shifted sensing}
\label{sec:sensing}

Monotone sensing maps first yield a maximum-principle property; a paired-sensitivity condition then yields velocity clusters over edges with infinite accumulated sensitivity, while uniform slope bounds guarantee global, exponential velocity alignment.

\subsection{Shifted Sensing System Description}\label{subsec:sensingdescription}

For shifted sensing, \eqref{equ_general_system} becomes
\begin{equation}\label{equ_relstate_system}
    \Sigma_i :  \begin{cases} 
        \dot{x}_i = u_i, \\
        y_i = x_i,
    \end{cases} \quad 
    \Psi_\ell : \,
        \tilde{z}_{\ell} = \shift_{\ell}(z_{\ell}).
\end{equation}

Consider an edge $\ell$ whose positive orientation is $\ki=(i,j)$. Agent $i$ measures $z_{\ki}=x_i-x_j=(B^\top x)_\ell$ and agent $j$ measures $z_{\ko}=x_j-x_i=-(B^\top x)_\ell$. We write $\shift_{ij}:=\shift_{\ki}$ and $\shift_{ji}:=\shift_{\ko}$ for the maps applied by agents $i$ and $j$, respectively. Hence, in general $\shift_{ij}(x_i-x_j) \neq -\shift_{ji}(x_j-x_i)$. By \eqref{equ_compact_control} and \eqref{equ_relstate_system},
\begin{equation}
\label{equ_shift_relative}
\dot{x} = -\Bo \shiftt^\mathrm{i} (B^\top x) +  \Bi \shiftt^\mathrm{o}(-B^\top x),
\end{equation}
where $\shiftt^\mathrm{i}(z) = [\shift_{1^\mathrm{i}}(z_1),\cdots, \shift_{M^\mathrm{i}}(z_M)]^\top$ collects the measurements along $\Ei$, and $\shiftt^\mathrm{o}$ is defined analogously for $\Eo$. Componentwise,
\begin{equation}\label{equ_shift_relative_component}
\dot{x}_i = - \sum_{j\in\mathcal{N}_i} \shift_{ij}(x_i-x_j),
\end{equation}
and we denote by $g:\mathbb{R}^N\to\mathbb{R}^N$ the right-hand side, $\dot x=g(x)$.

\subsection{Shifted Sensing Analysis}\label{subsec:sensinganalysis}

The core idea of the upcoming analysis relies on the maximum (minimum) principle, which is a tool in the study of partial differential equations, both in continuous settings \cite{evans2022partial} and in their discrete formulations \cite{strikwerda2004finite}. A similar approach has been used in opinion dynamics \cite{motsch2014heterophilious, vasca2021practical}. Its use in this paper allows us to show how the maximum and minimum velocities in the system approach each other during the evolution, and to concretely characterize this behavior through rigorous bounds. 

\begin{figure}[!htbp]
\centering
\includegraphics[width=\columnwidth]{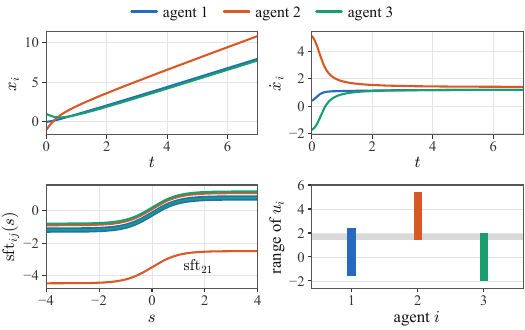}
\caption{Illustration of why shifted sensing requires stronger conditions than range compatibility alone. Three agents over a triangle graph. Reading row-wise, the four subfigures show the agent states, the state derivatives, the six scalar shifted sensing functions, colored by the agent applying them, and the ranges of the nodal inputs $u_i=-\sum_{j}\shift_{ij}(\cdot)$. Although these nodal ranges are compatible (gray band), the sensing functions follow $\shift_\ell(z_\ell)=\operatorname{tanh}(z_\ell)+b_\ell$ and do not satisfy the uniform sensitivity condition in Assumption~\ref{ass_Lipschitz}; hence global velocity alignment is not guaranteed, and in this case the velocities form two clusters, $\{1,3\}$ and $\{2\}$. The numerical trajectories exhibit a bounded $x_1-x_3$ and growing inter-cluster separations, consistent with the asymptotic sensitivity characterization in Theorem~\ref{thm_active_graph}.}
\label{fig: sftsen}
\end{figure}

While Lemma~\ref{lemma_onesol_nonginv} establishes existence and uniqueness of steady-state velocity under shifted actuation, analogous reasoning for shifted sensing is more complex. In shifted actuation, compatibility can be checked through the scalar intersection of nodal input ranges. In shifted sensing, the variables are edge states $z_\ell$ constrained by the cycle-compatible subspace $z\in\operatorname{im}(B^\top)$. Figure~\ref{fig: sftsen} illustrates this difference: compatible nodal ranges alone do not guarantee velocity alignment, because all agents must realize the same velocity at one compatible edge-state configuration.

The next result first establishes the maximum principle for monotone sensing. For differentiable maps, let $a_{ij}(t):=\shift'_{ij}(x_i(t)-x_j(t))$ denote the oriented sensitivity along an edge. We write $\mathcal G_a$ for the graph whose physical edges satisfy $\int_0^\infty a_{ij}(t)\,\mathrm dt=\infty$; the paired-sensitivity condition used below makes this definition independent of edge orientation. This graph identifies the interactions with infinite accumulated sensitivity.

\begin{theorem}\label{thm_active_graph}
\textit{Active-Graph Alignment -} Consider the shifted-sensing system~\eqref{equ_shift_relative} over a graph satisfying Assumption~\ref{ass_graph}, with maps satisfying Assumption~\ref{ass_shift}.
\begin{enumerate}[(i)]
\item Along every solution, the largest velocity $\overline v(t):=\max_i\dot x_i(t)$ is non-increasing and the smallest velocity $\underline v(t):=\min_i\dot x_i(t)$ is non-decreasing. Thus $S(t):=\overline v(t)-\underline v(t)$ is non-increasing.
\item Suppose in addition that the sensing maps are $C^1$ and satisfy the paired-sensitivity condition, for some $K\ge1$,
\begin{equation}\label{eq:paired_sensitivity}
K^{-1}\shift'_{ji}(-s)\le\shift'_{ij}(s)\le K\shift'_{ji}(-s),\quad s\in\mathbb R.
\end{equation}
Then every velocity converges to a finite limit $v_i^\infty$, with $v_i^\infty=v_j^\infty$ whenever $i$ and $j$ belong to the same connected component of $\mathcal G_a$.
\end{enumerate}
\end{theorem}
\begin{pf}
(i) We use a comparison argument that requires only the continuity
and monotonicity in Assumption~\ref{ass_shift}, not local Lipschitz
regularity.  Let $x(t)$ and $\widehat x(t)$ be any two solutions of
\eqref{equ_shift_relative}, and set $\delta_i(t)=x_i(t)-\widehat x_i(t)$.
For every node $p$ at which $\delta_p(t)=\max_i\delta_i(t)$,
$\delta_p-\delta_j\ge0$ for each neighbor $j$. Therefore
$x_p-x_j\ge\widehat x_p-\widehat x_j$, and monotonicity gives
\[
\dot\delta_p=-\sum_{j\in\mathcal N_p}
\big[\shift_{pj}(x_p-x_j)-\shift_{pj}(\widehat x_p-\widehat x_j)\big]
\le0.
\]
Similarly, $\dot\delta_q\ge0$ at every index $q$ attaining
$\min_i\delta_i$. As both solutions are continuously differentiable,
the upper Dini derivative of $\max_i\delta_i$ is non-positive,
and that of $\min_i\delta_i$ is non-negative. Consequently, the
largest difference between two solutions cannot increase, and the
smallest difference cannot decrease.

Now fix $\varepsilon>0$ and compare the solution $x(t)$ with its
time shift $x(t+\varepsilon)$, using
$\delta_i(t)=x_i(t+\varepsilon)-x_i(t)$. It follows that
\[
\max_i\frac{x_i(t+\varepsilon)-x_i(t)}{\varepsilon}
\quad\text{is non-increasing in }t,
\]
while the corresponding minimum is non-decreasing. Taking
$\varepsilon\to0^+$ and using continuity of $\dot x$ proves the
claim for $\overline v(t)$ and $\underline v(t)$. This also
implies that the velocities remain bounded by their initial
extrema; hence trajectories cannot have finite escape time.
The comparison applies to arbitrary solutions and also yields
uniqueness of solutions with the same initial state.

(ii) Differentiation along solutions gives
\[
\dot v_i=\sum_{j\in\mathcal N_i}a_{ij}(t)(v_j-v_i),\qquad a_{ij}(t)\ge0.
\]
Condition~\eqref{eq:paired_sensitivity} ensures $K^{-1}a_{ji}(t)\le a_{ij}(t)\le K a_{ji}(t)$ for every physical edge and time. Therefore this induced velocity system is cut-balanced. Applying Theorem~1(a)--(b) of~\cite{hendrickx2013cutbalanced} to this continuous-time velocity system yields existence of each $v_i^\infty$ and equality of limiting velocities on every connected component of the graph of infinite accumulated interactions, namely $\mathcal G_a$. \qed
\end{pf}

The graph $\mathcal G_a$ describes persistent sensitivity, not merely bounded relative distances. In particular, if $|x_i(t)-x_j(t)|$ remains bounded and $\shift'_{ij}$ is positive on that bounded interval, then $\{i,j\}$ belongs to $\mathcal G_a$. For $\shift_{ij}(s)=\tanh(s)+b_{ij}$, the paired sensitivities are identical ($K=1$); thus the two-cluster behavior in Fig.~\ref{fig: sftsen} is covered without requiring any finite limiting state. Condition~\eqref{eq:paired_sensitivity} is specific to the cluster-convergence claim: it need not hold when the two orientations have different decay rates (e.g., one uses $\tanh(s)$ and the other $s$), while non-$C^1$ sensing maps remain covered by part~(i). Global alignment is recovered under uniform sensitivity as follows.

\begin{assumption}\label{ass_Lipschitz}
\textit{Uniform Sensing Sensitivity -}
There exist constants $0<\klo\le\khi$ such that every oriented sensing function satisfies
\[
\klo\le\frac{\shift_{ij}(s)-\shift_{ij}(r)}{s-r}\le\khi\qquad\text{for all } s\neq r.
\]
\end{assumption}

For differentiable maps, Assumption~\ref{ass_Lipschitz} reads $\klo\le\shift'_{ij}\le\khi$. It is an all-edge-active condition: every sensing map is strictly increasing and onto $\mathbb R$, e.g., $\operatorname{tanh}(z-h)+cz+b$ with $c>0$, which excludes saturated or deadzone sensing, and $\klo$ prevents the state-dependent coupling weights from vanishing.

\begin{theorem}\label{thm_relative_state_general}
\textit{All-Edge Sensing Alignment -}
Consider the shifted-sensing closed-loop system~\eqref{equ_shift_relative} whose graph $\mathcal G$ satisfies Assumption~\ref{ass_graph}, and whose sensing shift functions satisfy Assumptions~\ref{ass_shift} and~\ref{ass_Lipschitz}. Then the system achieves global velocity alignment exponentially fast: $\lim_{t\to\infty}\dot x(t)=c\mathbf 1_N$ for some $c\in\mathbb R$, and, for any $T>0$, $S(t)\le(1-\gamma)^{\lfloor t/T\rfloor}S(0)$ with $\gamma\in(0,1)$ given in \eqref{eq:gamma}, depending only on $T$, $N$, the maximum graph degree, and the bounds $\klo,\khi$.
\end{theorem}

\begin{pf}
By Assumption~\ref{ass_shift}, the sensing maps are locally Lipschitz; since $x(t)$ is continuously differentiable, $v(t)=g(x(t))$ is locally Lipschitz on every compact time interval. Thus, for almost every $t$ we have $\frac{\mathrm d}{\mathrm dt}\shift_{ij}(z_{ij}(t))=\kappa_{ij}(t)\dot z_{ij}(t)$ with $\kappa_{ij}(t)\in[\klo,\khi]$. Indeed, if $\dot z_{ij}(t)\ne0$, the derivative is the limit of the product of the difference quotient of $\shift_{ij}$, which lies in $[\klo,\khi]$ by Assumption~\ref{ass_Lipschitz}, and the difference quotient of $z_{ij}$, which tends to $\dot z_{ij}(t)\neq0$; if $\dot z_{ij}(t)=0$, both sides vanish for any $\kappa_{ij}(t)$, e.g., $\kappa_{ij}(t)=\klo$. Hence, for almost every $t$,
\begin{equation}\label{eq:veldyn}
\dot v_i=-\sum_{j\in\mathcal N_i}\kappa_{ij}(t)\,(v_i-v_j),\qquad i\in\mathcal V,
\end{equation}
i.e., the velocities obey a consensus protocol over the original bidirectional graph with time-varying arc weights in $[\klo,\khi]$.

To estimate the contraction of the velocity spread, fix $t_0\ge0$ and $T>0$, let $\overline v_0:=\overline v(t_0)$, $S_0:=S(t_0)$, and define $e_i(t):=\overline v_0-v_i(t)$ for $t\ge t_0$. By Theorem~\ref{thm_active_graph}(i), $e_i(t)\ge0$. From \eqref{eq:veldyn}, $\dot e_i=\sum_{j\in\mathcal N_i}\kappa_{ij}(e_j-e_i)$, so that, with $a:=d_{\max}\khi$ and $d_{\max}$ the maximum degree,
\begin{equation}\label{eq:ecomp}
\dot e_i\ge-a\,e_i+\klo\, e_j\qquad\text{for every } j\in\mathcal N_i .
\end{equation}
Let $q$ satisfy $v_q(t_0)=\underline v(t_0)$, so $e_q(t_0)=S_0$, and let $\tau:=T/(N-1)$. Since $\dot e_q\ge-ae_q$, we have $e_q(t)\ge\gamma_0S_0$ on $[t_0,t_0+T]$ with $\gamma_0:=e^{-aT}$. Suppose that an agent $j$ satisfies $e_j(s)\ge\gamma_{k-1}S_0$ for $s\in[t_0+(k-1)\tau,t_0+T]$, and let $i\in\mathcal N_j$. Multiplying \eqref{eq:ecomp} by $e^{a t}$ and integrating from $s_0:=t_0+(k-1)\tau$, and using $e_i\ge0$, yields for $t\in[t_0+k\tau,t_0+T]$
\[
e_i(t)\ge\klo\gamma_{k-1}S_0\int_{s_0}^{t}e^{-a(t-s)}\mathrm ds\ge\gamma_kS_0,
\]
with $\gamma_k:=\klo\tau e^{-aT}\gamma_{k-1}$.
Since $\mathcal G$ is connected, every agent is at graph distance at most $N-1$ from $q$, so by induction all agents satisfy $e_i(t_0+T)\ge\gamma S_0$ with
\begin{equation}\label{eq:gamma}
\gamma:=\min_{0\le k\le N-1}\gamma_k,\quad \gamma_k=e^{-aT}\big(\klo\tau e^{-aT}\big)^{k}.
\end{equation}
Therefore $\overline v(t_0+T)=\overline v_0-\min_ie_i(t_0+T)\le\overline v_0-\gamma S_0$, and since $\underline v(t_0+T)\ge\underline v(t_0)$ by Theorem~\ref{thm_active_graph}(i), $S(t_0+T)\le(1-\gamma)S_0$. Because $\gamma$ does not depend on $t_0$, iterating gives $S(t)\le(1-\gamma)^{\lfloor t/T\rfloor}S(0)$. Finally, $\overline v$ is non-increasing, $\underline v$ is non-decreasing, and their difference vanishes, so both converge to a common limit $c$, i.e., $\dot x(t)\to c\mathbf 1_N$. \qed
\end{pf}

The factor $\gamma$ in \eqref{eq:gamma} is conservative: it certifies an exponential rate rather than estimating it.

\begin{remark}
\textit{Sensing Compatibility -}
Theorem~\ref{thm_active_graph} does not determine the clusters a priori, and their velocities are not given by a scalar intersection of nodal ranges. Writing $g(x)=\hat g(B^\top x)$ with $\hat g(z):=-\Bo\shiftt^{\mathrm i}(z)+\Bi\shiftt^{\mathrm o}(-z)$, a coherent constant-velocity motion with fixed finite relative states exists if and only if there are $z\in\operatorname{im}(B^\top)$ and $c\in\mathbb R$ satisfying $\hat g(z)=c\mathbf 1_N$. This feasibility condition by itself does not establish attraction to that motion. This is the sensing-side analogue of the range-intersection condition for shifted actuation in Assumption~\ref{ass_range}, but it is more involved because the feasible variable is constrained by the cycle-compatible edge subspace $\operatorname{im}(B^\top)$.
\end{remark}

\begin{lemma} \label{LaSalleThm1}
\textit{Global LaSalle Criterion (\cite{khalil2002nonlinear}, Corollary 4.2)}
Let $x=0$ be an equilibrium point of an autonomous first-order system with locally Lipschitz right-hand side. Let $V:\mathbb{R}^N\rightarrow\mathbb{R}$ be a continuously differentiable, radially unbounded, positive definite function such that $\dot{V}(x)\leq 0$ for all $x\in\mathbb{R}^N$. Let $\mathcal{Z}=\{x\in\mathbb{R}^N|\dot{V}(x)=0\}$ and suppose that no solution can stay identically in $\mathcal{Z}$ other than the trivial solution $x(t)\equiv 0$. Then, the origin is globally asymptotically stable.
\end{lemma}
\begin{proposition}\label{prop_odd}
\textit{Odd Sensing Consensus -}
Consider the shifted-sensing closed-loop system \eqref{equ_shift_relative} under Assumptions~\ref{ass_graph} and \ref{ass_shift}. Suppose all shift functions are odd and satisfy $s\shift_{\ki}(s)>0$ for every $s\neq0$ and every positive orientation $\ki\in\Ei$. For each edge $\ell\in\mathcal{E}$, let its two oriented measurements be $\ki\in\Ei$ and $\ko\in\Eo$, and suppose further that $\shift_{\ki}(\cdot)=\shift_{\ko}(\cdot)$. Under these assumptions, the system achieves a static consensus configuration.
\end{proposition}

\begin{pf}
For identical odd sensing maps in both orientations, the global LaSalle criterion in Lemma~\ref{LaSalleThm1} applies to an edge-potential construction. Let $z=B^\top x$. Under the oddness and edge-wise equality assumptions, \eqref{equ_shift_relative} reduces to $\dot x=-B\shiftt^{\mathrm{i}}(z)$. Hence the edge dynamics are
\begin{equation}
    \dot{z} = - B^\top B\shiftt^{\mathrm{i}}(z) = -L_e\shiftt^{\mathrm{i}}(z),
\end{equation}
which evolve in $\operatorname{im}(B^\top)$. Consider the following Lyapunov-candidate function $V:\operatorname{im}(B^\top)\to \mathbb{R}$
\begin{equation}
V(z) = \sum_{\ki\in\Ei} \int_{0}^{z_{\ki}} \shift_{\ki}(s) \mathrm{d}s, 
\end{equation}
where the sum is performed over the positive edges of the graph. By the strict sign condition, $s\,\shift_{\ki}(s)>0$ for $s\neq0$, so each term is positive for $z_{\ki}\neq0$ and, for $|z_{\ki}|\ge1$, satisfies $\int_0^{z_{\ki}}\shift_{\ki}(s)\mathrm ds\ge c_{\ki}|z_{\ki}|$, where $c_{\ki}:=\frac12\min\{\shift_{\ki}(\tfrac12),-\shift_{\ki}(-\tfrac12)\}>0$, using the positive and negative half-axes separately. Hence $V$ is positive definite and radially unbounded. Moreover, it satisfies $\dot{V} = \sum_{\ki\in\Ei} \shift_{\ki}(z_{\ki}) \dot{z}_{\ki} = (\shiftt^{\mathrm{i}}(z))^\top \dot{z}$. Therefore, $\dot{V} = -(\shiftt^{\mathrm{i}}(z))^\top B^\top B \shiftt^{\mathrm{i}}(z) = -\|B\shiftt^{\mathrm{i}}(z)\|^2 = -\|\dot x\|^2$. In order to apply LaSalle's Invariance Principle (Lemma~\ref{LaSalleThm1}), we study the set $\mathcal{Z} = \{z\in\operatorname{im}(B^\top) \, | \, \dot V = 0\}$. Then, for all $z\in \mathcal{Z}$ it must be that $B \shiftt^{\mathrm{i}}(z)=\mathbf{0}_N$. We claim this can only happen if $z=\mathbf{0}_M$. In fact, since $z\in\operatorname{im}(B^\top)$, there exists some $x\in\mathbb{R}^N$ such that $z=B^\top x$; hence, for $z\in \mathcal{Z}$, it must be that $z^\top \shiftt^{\mathrm{i}}(z) = x^\top B\shiftt^{\mathrm{i}}(z) = 0$. Simultaneously, $z^\top \shiftt^{\mathrm{i}}(z) = \sum_{\ki \in \Ei} z_{\ki} \shift_{\ki}(z_{\ki})$. By the strict sign condition, if a relative state is nonzero, $z_{\ki}\ne0$, its corresponding term in the sum is strictly positive; hence all relative states are zero, $z_{\ki}=0$. Therefore, the set $\mathcal{Z}$ contains a single point $\mathcal{Z} = \{ \mathbf{0}_M \}$, which is an equilibrium point of the dynamical system. Since $\operatorname{im}(B^\top)$ is invariant and isomorphic to $\mathbb R^{N-1}$, the LaSalle criterion applies on this subspace. Application of Lemma~\ref{LaSalleThm1} shows that the system converges globally asymptotically to $z=\mathbf{0}_M$, corresponding to a static consensus equilibrium. \qed
\end{pf}

The strict sign condition is essential for Proposition~\ref{prop_odd}: with the odd deadzone map $\shift(z)=0$ for $|z|\le1$ and $\shift(z)=z-\operatorname{sign}(z)$ otherwise, every non-consensus configuration with $|x_i-x_j|\le1$ on all edges is a static equilibrium. The proof only uses the pairwise antisymmetry $\shift_{\ko}(-s)=-\shift_{\ki}(s)$, which may replace oddness and the edge-wise equality. Such homogeneous endpoint measurements are nongeneric; in general, relative-state inconsistencies propagate and drifts occur.

\section{Linear shift functions}
\label{sec:linear}
The linear case admits an explicit solution and approximates the unsaturated regime. With $\shift(s) = as + b$, $a>0$, with the appropriate node or oriented-edge sub-index, horizontal and vertical shifts produce identical dynamics, since $a(s-h)+b = as+(b-ah)$, unlike the nonlinear case~(Subsection~\ref{subsec:role}).

Both \eqref{equ_input_system} and \eqref{equ_relstate_system} lead to
\begin{equation}\label{equ_dyn_both_linear_sh}
\dot{x} = - \tilde{L} x + b,
\end{equation}
where for the shifted actuation case we have $\tilde{L} = \operatorname{diag}(a_i)L$ and $b=[b_1,\cdots,b_N]^\top$ with positive slopes $a_i>0$, $i\in\mathcal{V}$. In the shifted sensing case, consistent with \eqref{equ_shift_relative},
\[
\tilde{L}=B^\mathrm{o}D^\mathrm{i}B^\top+B^\mathrm{i}D^\mathrm{o}B^\top,
\qquad
b=-B^\mathrm{o}b^\mathrm{i}+B^\mathrm{i}b^\mathrm{o},
\]
where $D^\mathrm{i}=\operatorname{diag}(a_{\ki})$ and $D^\mathrm{o}=\operatorname{diag}(a_{\ko})$ contain the positive slopes $a_{\ki},a_{\ko}>0$, while $b^\mathrm{i}$ and $b^\mathrm{o}$ collect the initial-side and terminal-side bias terms, respectively. In either case, $\tilde{L}$ is a weighted Laplacian (zero row sums, non-positive off-diagonal entries) whose kernel is spanned by $\mathbf{1}_N$, all other eigenvalues having positive real part \cite{ren2005consensus}, and its left null vector $v$ can be chosen positive, so $v^\top\mathbf{1}_N>0$ \cite{bullo2018lectures}.

Both linear interaction channels admit \eqref{equ_dyn_both_linear_sh}, and Proposition~\ref{prop_linear} applies to both. Every linear shifted actuation is reproduced by a linear shifted sensing with slopes $a_\ell=a_i$ and biases $b_\ell=-b_i/|\mathcal{E}_i|$ for $\ell\in\mathcal{E}_i$. Conversely, since nodal biases can always be distributed in this way, a linear shifted sensing is a linear shifted actuation if and only if each agent applies a common slope to all its measurements, i.e., $a_\ell=a_i$ for all $\ell\in\mathcal{E}_i$. Hence linear actuation is a special case of linear sensing; e.g., $\operatorname{diag}(a_i)L$ has a real spectrum, being similar to $\operatorname{diag}(a_i)^{1/2}L\operatorname{diag}(a_i)^{1/2}$, while sensing Laplacians may have complex eigenvalues.

\begin{proposition}\label{prop_linear}
\textit{Linear Drift Formula -}
Consider the multi-agent system \eqref{equ_dyn_both_linear_sh} under Assumption~\ref{ass_graph}, corresponding to using linear shift functions for $F_i$ and $\psi_{ \ell}$ in \eqref{equ_general_system}. Then
\[
x(t)=P_1bt+\tilde{L}^{\dagger}P_2b+c'\mathbf{1}_N+o(1),\qquad t\to\infty,
\]
where $P_1=\frac{\mathbf{1}_Nv^\top}{\mathbf{1}_N^\top v}$, $v$ is the left eigenvector of $\tilde{L}$ associated with the zero eigenvalue, $P_2=I-P_1$, and $c'=\frac{v^\top(x(0)-\tilde{L}^{\dagger}P_2b)}{v^\top\mathbf{1}_N}$. The convergence is exponential.
\end{proposition}

\begin{pf}
Note that $P_1$ and $P_2$ are complementary projections, with $\tilde{L}P_1=0$ and $v^\top P_2=0$. Let $c:=\frac{v^\top b}{v^\top\mathbf{1}_N}$, so that $P_1b=c\mathbf{1}_N$. Since $v^\top P_2 b=0$ and $\operatorname{im}(\tilde L)=\ker(v^\top)$, the vector $P_2b$ lies in $\operatorname{im}(\tilde L)$; as $\tilde L\tilde L^\dagger$ is the orthogonal projection onto $\operatorname{im}(\tilde L)$, the vector $x^\ast:=\tilde{L}^{\dagger}P_2b+c'\mathbf{1}_N$ satisfies $\tilde Lx^\ast=P_2b$. Define $\xi(t):=x(t)-c\mathbf{1}_Nt-x^\ast$. Using \eqref{equ_dyn_both_linear_sh},
\[
\begin{aligned}\dot\xi&=-\tilde L(\xi+c\mathbf{1}_Nt+x^\ast)+b-c\mathbf{1}_N\\&=-\tilde L\xi+b-P_2b-P_1b=-\tilde L\xi.\end{aligned}
\]
Moreover, $v^\top\dot x=v^\top b$ gives $v^\top x(t)=v^\top x(0)+(v^\top b)t$, and the choice of $c'$ gives $v^\top x^\ast=v^\top x(0)$; hence $v^\top\xi(t)=0$ for all $t$. The subspace $\ker(v^\top)=\operatorname{im}(\tilde L)$ is invariant under $\tilde L$ and complementary to $\operatorname{span}(\mathbf 1_N)$, because $v^\top\mathbf 1_N\neq0$, so the restriction of $\tilde L$ to it has exactly the nonzero eigenvalues of $\tilde L$, all with positive real part. Therefore $\xi(t)\to0$ exponentially, which proves the claim with $\lim_{t\to\infty}\dot x(t)=c\mathbf{1}_N=P_1b$. The constant vector $x^\ast$ determines the eventual relative states, which are not zero for heterogeneous shift functions in general. \qed
\end{pf}

The bias splits into $b_\parallel := P_1 b$, which determines the final velocity, and $b_\perp:=P_2 b$, which determines the \emph{non-consensus distortion} $\tilde{L}^{\dagger} b_\perp$. The norm of the distortion is bounded by $\| \tilde{L}^{\dagger} b_\perp \|_2 \le \| \tilde{L}^{\dagger} \|_2 \,\| b_\perp \|_2 = \frac{1}{\sigma_2(\tilde{L})}\, \| b_\perp \|_2$, where $\sigma_2(\tilde{L})$ is the smallest nonzero singular value of $\tilde L$.

\begin{remark}
\textit{Connectivity Distortion Bound -} 
If $\tilde{L}$ is symmetric, then $\sigma_2(\tilde{L})=\lambda_2(\tilde{L})$ is the algebraic connectivity, and the distortion bound reduces to $1/\lambda_2$. This holds, e.g., for identical actuation slopes, or for $a_{\ki}=a_{\ko}$ on every edge; then better connectivity reduces the maximum distortion.
\end{remark}

\section{Simulations}
\label{sec:simulation}
Three scenarios on the graph of Figure~\ref{fig:graph} illustrate the results: non-negative actuation with flat regions, as in systems without reverse thrust or with jammed actuators; smooth actuator calibration maps, whose drift and limiting offsets are predicted in closed form; and biased nonlinear relative sensing. We close with second-order agents.

\begin{figure}[!htbp]
\centering
\includegraphics[width=\columnwidth]{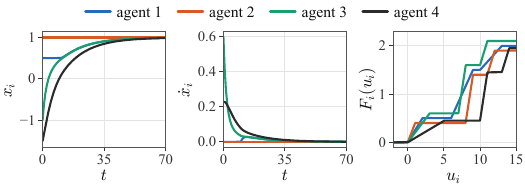}
\caption{Piecewise linear non-negative shifted actuation in system~\eqref{equ_input_system}. From left to right, the three subfigures show the agent states, the state derivatives, and the non-negative shifted actuation functions. The agents reach static consensus at the maximum initial state.}
\label{fig: sim1}
\end{figure}

Figure~\ref{fig: sim1} illustrates the first example, which uses piecewise linear non-negative actuation. Although these functions are not invertible, their ranges have a common intersection containing the origin, so the range-compatibility condition in Assumption~\ref{ass_range} holds and Lemma~\ref{lemma_onesol_nonginv} gives $y^*=0$. Moreover, every $F_i$ vanishes exactly on $u\le0$ and is positive for $u>0$, so $-F_i^{-1}(0)=[0,\infty)$ and the set $E$ in the proof of Theorem~\ref{thm_inp_constrained2} reduces to $\{\eta:\sum_i\eta_i=0,\ \eta_i\ge0\}=\{\mathbf 0_N\}$. Hence static consensus is guaranteed by the theory, and the agent initialized at the maximum state remains stationary while the others move toward it.

\begin{figure}[!htbp]
\centering
\includegraphics[width=\columnwidth]{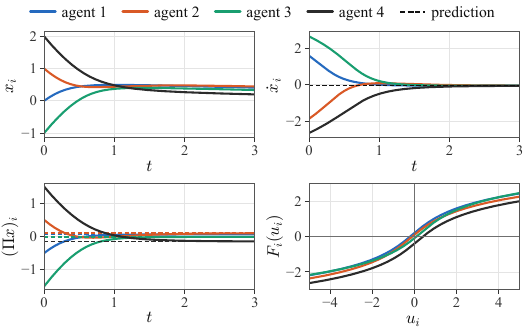}
\caption{Smooth invertible shifted actuation with $F_i(u)=\operatorname{asinh}(u-h_i)+b_i$. Reading row-wise, the four subfigures show the agent states, the state derivatives, the non-consensus components of the state, and the invertible shifted actuation functions. The dashed curves in the third subfigure denote the closed-form predictions for the limiting non-consensus components, and the dashed horizontal line in the second subfigure denotes the aligned velocity.}
\label{fig: sim2}
\end{figure}

The second example uses the invertible maps of Figure~\ref{fig: sim2}, which satisfy Assumption~\ref{ass_ginv}. All velocities converge to the same nonzero value, i.e., a drifting rather than static configuration. The dashed reference lines, computed from $y^\ast$ and $L^\dagger\bar\eta$ as in Section~\ref{sec:actuation}, match the limits of the simulated trajectories.

\begin{figure}[!htbp]
\centering
\includegraphics[width=\columnwidth]{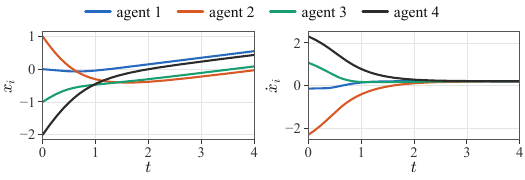}
\caption{Nonlinear shifted sensing in system~\eqref{equ_relstate_system} with $\shift_\ell(z)=\operatorname{tanh}(z-h_\ell)+0.2z+b_\ell$. The left subfigure shows the agent states and the right subfigure shows the state derivatives. Velocity alignment is achieved under Assumption~\ref{ass_Lipschitz}.}
\label{fig: sim3}
\end{figure}

For comparison, the nonlinear sensing maps in Figure~\ref{fig: sim3} satisfy Assumption~\ref{ass_Lipschitz} with $\klo=0.2$ and $\khi=1.2$. Their states do not approach static values, whereas the velocities align, in agreement with Theorem~\ref{thm_relative_state_general}.

To illustrate Remark~\ref{rem_higher}, consider double integrators with linear actuation shifts and the standard second-order protocol, $\ddot x=\operatorname{diag}(a_i)u+b$ with $u=-L(x+\gamma\dot x)$, $a_i>0$, and $\gamma>0$. Since $\operatorname{diag}(a_i)L$ is similar to $\operatorname{diag}(a_i)^{1/2}L\operatorname{diag}(a_i)^{1/2}$, each disagreement mode satisfies $s^2+\gamma\lambda s+\lambda=0$ for an eigenvalue $\lambda>0$ of $\operatorname{diag}(a_i)L$, and is exponentially stable. The left null vector of $\operatorname{diag}(a_i)L$ is $w=\operatorname{diag}(a_i)^{-1}\mathbf 1_N$, so $w^\top\ddot x=w^\top b$, and the argument of Proposition~\ref{prop_linear} gives $\ddot x(t)\to c\mathbf 1_N$ with $c=\sum_ib_i/a_i\,/\sum_i1/a_i$: the shifts produce a common constant acceleration instead of a common velocity. On the graph of Figure~\ref{fig:graph} with $a=(1,0.8,1.2,1)$, $b=(0.2,-0.1,0.3,0)$, and $\gamma=1$, this predicts $c\approx0.0796$, which the accelerations in Figure~\ref{fig: sim4} match to within $10^{-9}$; consequently, the velocities grow linearly. If instead $\sum_ib_i/a_i=0$, e.g., $b=(0.8,-0.96,1.2,-0.6)$, then $w^\top\dot x$ is constant and agents starting at rest reach a static configuration with offsets $L^\dagger\operatorname{diag}(a_i)^{-1}b$.

{
\begin{figure}[!htbp]
\centering
\includegraphics[width=\columnwidth]{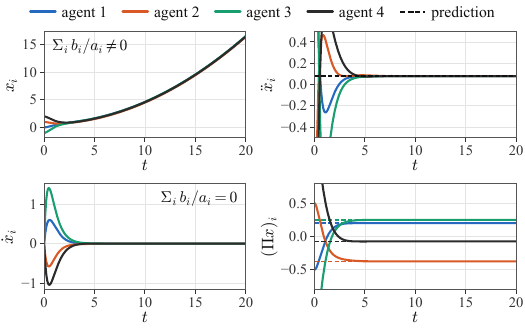}
\caption{Second-order agents with linear actuation shifts. The first two subfigures show the agent states and accelerations for $b=(0.2,-0.1,0.3,0)$, with $\sum_ib_i/a_i\neq0$: the accelerations align at the predicted value $c$ (dashed line), so the velocities grow linearly. The last two subfigures show the velocities and the non-consensus components of the state for $b=(0.8,-0.96,1.2,-0.6)$, with $\sum_ib_i/a_i=0$: the agents reach a static configuration with the predicted offsets (dashed lines).}
\label{fig: sim4}
\end{figure}
}

\section{Conclusion}
\label{sec:conclusion}
This paper has represented actuator saturation, unilateral actuation, biased measurements, and asymmetric edge interactions as horizontal and vertical shifts of monotone functions, which allows comparing imperfections in the actuation and sensing channels within the same consensus model. For shifted actuation, the analysis gives a range-based check for velocity alignment and includes bounded as well as non-invertible maps. For shifted sensing, the maximum-principle analysis first yields non-increasing velocity spread and cluster convergence under a paired-sensitivity condition, without assuming state convergence, and then recovers global exponential velocity alignment when all sensing edges remain responsive.

The results show that the location and shape of the shift are both essential. Linear shifted interactions lead to the same class of affine closed-loop descriptions for actuation and sensing, actuation being a special case of sensing, whereas nonlinear shifted interactions may generate channel-dependent steady-state offsets, static consensus, or drifting consensus. In practice, actuator ranges indicate whether a common velocity can be realized, sensing responsiveness whether the active graph remains connected, and linear calibration maps anticipate the drift and the limiting configuration. Future work will relax the all-edge-active condition for shifted sensing, especially for saturated and deadzone-type measurement maps, and will extend the proposed framework to higher-order dynamics, coupled multidimensional shifts, and one-way or time-varying interaction graphs.

\appendices

\section{Proof of Lemma~\ref{lemma_onesol_nonginv}}
\label{proof: lemma_onesol_nonginv}
\begin{pf}
For existence, we seek a common velocity $y\in\mathcal R$ and numbers
$\eta_i\in-\shift_i^{-1}(y)$ satisfying $\sum_i\eta_i=0$.  Equivalently,
put $u_i=-\eta_i$ and $H:=\{u\in\mathbb R^N:\mathbf1_N^\top u=0\}$.
Consider the continuously differentiable, convex function
\[
 W(u)=\sum_{i\in\mathcal V}\int_0^{u_i}\shift_i(s)\,\mathrm ds,
 \qquad u\in H.
\]
Let $a<b$ lie in the interior of the common range $\mathcal R$ and
write $c=(a+b)/2$.  For each $i$, choose $r_i^-,r_i^+\in\mathbb R$
with $\shift_i(r_i^-)=a$ and $\shift_i(r_i^+)=b$.
Monotonicity of $\shift_i$ gives, for a suitable finite constant $C_i$,
\begin{equation}\label{eq:appendix-coercive}
 \int_0^{u_i}\shift_i(s)\,\mathrm ds
 \ge c u_i+\frac{b-a}{2}|u_i|-C_i,
 \qquad u_i\in\mathbb R.
\end{equation}
Indeed, for $u_i\ge r_i^+$ the integrand is at least $b$, whereas
for $u_i\le r_i^-$ it is at most $a$; on the intervening compact
interval the difference between the two sides has a finite lower bound.
Since $\sum_i u_i=0$, summing \eqref{eq:appendix-coercive} yields
\[
 W(u)\ge\frac{b-a}{2}\|u\|_1-\sum_i C_i,\qquad u\in H.
\]
Thus $W$ is coercive on the closed subspace $H$ and admits a minimizer
$\bar u\in H$.  The first-order condition is
$\nabla W(\bar u)\in H^\perp=\operatorname{span}\{\mathbf1_N\}$,
that is, $\shift_i(\bar u_i)=y^\ast$ for every $i$, for some
$y^\ast\in\mathcal R$.  Setting $\bar\eta=-\bar u$ supplies the required
zero-sum configuration.  Notice that this coercivity estimate is
independent of strict monotonicity and of whether the shift maps have
finite saturation bounds.  In particular, the same argument applies to
the Lyapunov potential $V(\eta)=W(-\eta)$ appearing in
Theorem~\ref{thm_inp_constrained2}, with $\eta\in\operatorname{im}(L)$.
The corresponding minimizer set is therefore nonempty and compact,
even when different input values yield the same common velocity.
This observation separates the existence of a compatible input vector
from the uniqueness of its associated output value.

For uniqueness of the compatible velocity,
Suppose $y_1<y_2$ both admitted such a configuration, and let
$u_i^{(r)}\in\shift_i^{-1}(y_r)$ with $\sum_i u_i^{(r)}=0$,
$r=1,2$.  Because each $\shift_i$ is non-decreasing, every preimage of
$y_1$ is strictly smaller than every preimage of $y_2$.
Consequently $u_i^{(1)}<u_i^{(2)}$ for all $i$, contradicting both
zero-sum equalities. Hence $y^\ast$ is unique. \qed
\end{pf}

\begin{figure}[!htbp]
    \centering
    \includegraphics[width=\columnwidth]{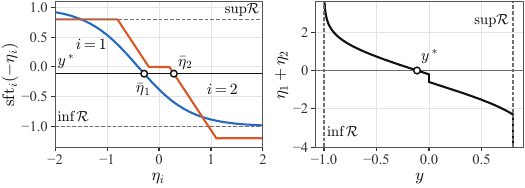}
    \caption{Illustration used in Lemma~\ref{lemma_onesol_nonginv}. The left subfigure shows two shifted functions, evaluated at $-\eta_i$, and the aligned velocity value $y^\ast$ for which the corresponding variables $\eta_i$, $i=1,2$, sum to zero. The right subfigure shows the possible values of $\sum_{i\in\mathcal V}\eta_i$ for each aligned velocity $y=\shift_i(-\eta_i)$ and their monotone variation with respect to $y$.}
    \label{fig:intersection}
\end{figure}

The geometric interpretation in Fig.~\ref{fig:intersection} uses the
preimage intervals
$T_i(y):=\{\eta_i:\shift_i(-\eta_i)=y\}$ and their Minkowski sum
$T(y):=\sum_iT_i(y)$ for $y\in\mathcal R$.
Each $T_i(y)$ is an interval because $\shift_i$ is continuous and
non-decreasing.  For $y_1<y_2$, every element of $T_i(y_1)$ exceeds
every element of $T_i(y_2)$; hence the admissible sums move in the
opposite direction as the common velocity increases.
The proof of Lemma~\ref{lemma_onesol_nonginv} establishes that precisely
one $y^\ast\in\mathcal R$ admits $0\in T(y^\ast)$, without assuming
bounded inverses or continuity of preimage endpoints at saturation.
Figure~\ref{fig:intersection} illustrates this compatibility relation rather than supplying
the existence argument. The left panel shows the two-agent case;
the right panel depicts the corresponding interval sums as the
candidate common velocity varies.  Flat intervals in the shifted maps
can make the compatible vector $\bar\eta$ non-unique, even though its
associated velocity $y^\ast$ is unique.

Flat actuator regions may admit several zero-sum compatible input vectors at the same velocity $y^\ast$. Theorem~\ref{thm_inp_constrained2} nevertheless guarantees convergence of each trajectory to one such vector, potentially depending on its initial condition. Under global invertibility the compatible vector is unique and is determined by the shift maps.

\section{Proof of Lemma~\ref{lemma_convexity}}
\label{proof: lyaconvex}
\begin{pf}
Let $\nabla f(x) = \operatorname{col}(h_1(x_1),\dots,h_k(x_k))$ with each $h_i$ non-decreasing, and fix $x,y\in\mathbb{R}^k$. Integrating $\frac{\mathrm d}{\mathrm dt}f(x+t(y-x))$ over $[0,1]$ gives
$f(y)-f(x)-\nabla f(x)^\top (y-x)=\int_0^1 \sum_{i=1}^k \big(h_i(x_i+t(y_i-x_i)) - h_i(x_i)\big)(y_i-x_i)\,\mathrm{d}t\ge0$, since each summand is non-negative by monotonicity of $h_i$. This is the first-order characterization of convexity. \qed
\end{pf}

\section{Proof of Lemma~\ref{lemma_radially}}
\label{proof: lyaunbounded}
\begin{pf}
Since $f$ is convex and differentiable, any stationary point is a global minimizer; hence $\mathcal X^\star=\arg\min f$, and we let $f^\star:=f(\xi^\star)$ for $\xi^\star\in\mathcal X^\star$.

{To establish radial unboundedness, argue by contradiction. Assume $f$ is \emph{not} radially unbounded. Then there exists some $c>f^\star$ such that the sublevel set $\Lambda_c:=\{\xi\in\mathbb{R}^k:\; f(\xi)\le c\}$ is unbounded. The set $\Lambda_c$ is closed and convex, and every unbounded closed convex set contains a ray. Indeed, fix $\alpha\in\Lambda_c$ and take $\xi_n\in\Lambda_c$ with $\|\xi_n\|\to\infty$. Up to a subsequence, $(\xi_n-\alpha)/\|\xi_n-\alpha\|\to v$ with $\|v\|=1$. For each $t\ge0$ and $n$ large, $\alpha+t(\xi_n-\alpha)/\|\xi_n-\alpha\|\in\Lambda_c$ by convexity, so $\alpha+tv\in\Lambda_c$ by closedness. Now fix $\xi^\star\in\mathcal X^\star$, $t\ge0$, and $\lambda\in(0,1]$. By convexity,
\[
f\big((1-\lambda)\xi^\star+\lambda(\alpha+\tfrac{t}{\lambda}v)\big)\le(1-\lambda)f^\star+\lambda c .
\]
The argument on the left equals $(1-\lambda)\xi^\star+\lambda\alpha+tv$, which tends to $\xi^\star+tv$ as $\lambda\to0$, while the right-hand side tends to $f^\star$. By continuity of $f$, $f(\xi^\star+tv)\le f^\star$, so $\xi^\star+tv\in\mathcal X^\star$ for every $t\ge0$. This contradicts the boundedness of $\mathcal X^\star$. Therefore, every sublevel set of $f$ is bounded, which is equivalent to $f(\xi)\to+\infty$ as $\|\xi\|\to\infty$. In other words, $f$ is radially unbounded. \qed}
\end{pf}

\bibliographystyle{IEEEtran}
\bibliography{refs}

@Article{de2021maneuvering,
  author    = {de Marina, Hector Garcia},
  journal   = {IEEE Transactions on Automatic Control},
  title     = {Maneuvering and robustness issues in undirected displacement-consensus-based formation control},
  year      = {2021},
  number    = {7},
  pages     = {3370--3377},
  volume    = {66},
  publisher = {IEEE},
}

@book{bullo2018lectures,
  title={Lectures on network systems},
  author={Bullo, Francesco and Cort{\'e}s, Jorge and D{\"o}rfler, Florian and Mart{\'\i}nez, Sonia},
  volume={1},
  year={2018},
  publisher={CreateSpace PlatformScotts Valley}
}

@article{chen2019convergence,
  title={Convergence analysis of signed nonlinear networks},
  author={Chen, Hao and Zelazo, Daniel and Wang, Xiangke and Shen, Lincheng},
  journal={IEEE Transactions on Control of Network Systems},
  volume={7},
  number={1},
  pages={189--200},
  year={2019},
  publisher={IEEE}
}

@inproceedings{xu2025distributed,
  title     = {Distributed Oscillatory Guidance for Formation Flight of Fixed-Wing Drones},
  author    = {Xu, Yang and Bautista, Jesus and Hinojosa, Jose and Garcia de Marina, Hector},
  booktitle = {2025 IEEE/RSJ International Conference on Intelligent Robots and Systems (IROS)},
  pages     = {2584--2591},
  year      = {2025},
  doi       = {10.1109/IROS60139.2025.11246740},
  organization = {IEEE}
}

@article{yang2014global,
  title={Global consensus for discrete-time multi-agent systems with input saturation constraints},
  author={Yang, Tao and Meng, Ziyang and Dimarogonas, Dimos V and Johansson, Karl H},
  journal={Automatica},
  volume={50},
  number={2},
  pages={499--506},
  year={2014},
  publisher={Elsevier}
}

@article{lim2015consensus,
  title={Consensus under saturation constraints in interconnection states},
  author={Lim, Young-Hun and Ahn, Hyo-Sung},
  journal={IEEE Transactions on Automatic Control},
  volume={60},
  number={11},
  pages={3053--3058},
  year={2015},
  publisher={IEEE}
}

@article{zou2023consensus,
  title={Consensus of saturated multiagent systems: {Tolerance} to nonidentical asymmetric saturation levels},
  author={Zou, Yao and Zuo, Zongyu and Xia, Kewei and Ding, Zhengtao},
  journal={IEEE Transactions on Automatic Control},
  volume={69},
  number={3},
  pages={1804--1811},
  year={2024},
  doi={10.1109/TAC.2023.3326684},
  publisher={IEEE}
}

@article{wang2024consensus,
  title={Consensus algorithms for double-integrator dynamics with different velocity and actuator saturation constraints},
  author={Wang, Gang and Zuo, Zongyu},
  journal={Automatica},
  volume={170},
  pages={111860},
  year={2024},
  publisher={Elsevier}
}

@article{zhu2025distributed,
  author    = {Zhu, Haitao and Lu, Jianquan and Lou, Yijun and Yang, Xinsong},
  title     = {Distributed Saturated Impulsive Quasi-Consensus for Leader-Follower Multi-Agent Systems: An Open Topology Framework},
  journal   = {IEEE/CAA Journal of Automatica Sinica},
  volume    = {12},
  number    = {9},
  pages     = {1941--1943},
  year      = {2025},
  doi       = {10.1109/JAS.2024.124743},
  publisher = {IEEE}
}

@article{fu2023safe,
  title={Safe consensus tracking with guaranteed full state and input constraints: A control barrier function-based approach},
  author={Fu, Junjie and Wen, Guanghui and Yu, Xinghuo},
  journal={IEEE Transactions on Automatic Control},
  volume={68},
  number={12},
  pages={8075--8081},
  year={2023},
  publisher={IEEE}
}

@article{lin2023constrained,
  title={Constrained containment control for multiagent systems with possible empty intersection of position constraints},
  author={Lin, Peng and Wang, Yaozhong and Wang, Dong and Wu, Zheng-Guang},
  journal={IEEE Transactions on Automatic Control},
  volume={69},
  number={3},
  pages={1842--1849},
  year={2024},
  doi={10.1109/TAC.2023.3332548},
  publisher={IEEE}
}

@article{wang2023connectivity,
  title={Connectivity-maintaining consensus of multi-agent systems with communication management based on predictive control strategy},
  author={Wang, Jie and Li, Shaoyuan and Zou, Yuanyuan},
  journal={IEEE/CAA Journal of Automatica Sinica},
  volume={10},
  number={3},
  pages={700--710},
  year={2023},
  publisher={IEEE}
}

@article{zhang2015satellite,
  title={Satellite cluster flight using on-off cyclic control},
  author={Zhang, Hao and Gurfil, Pini},
  journal={Acta Astronautica},
  volume={106},
  pages={1--12},
  year={2015},
  publisher={Elsevier}
}

@book{khalil2002nonlinear,
  title={Nonlinear systems},
  author={Khalil, Hassan K and Grizzle, Jessy W},
  volume={3},
  year={2002},
  publisher={Prentice hall Upper Saddle River, NJ}
}

@ARTICLE{ren2005consensus,
  author={Wei Ren and Beard, R.W.},
  journal={IEEE Transactions on Automatic Control}, 
  title={Consensus seeking in multiagent systems under dynamically changing interaction topologies}, 
  year={2005},
  volume={50},
  number={5},
  pages={655-661},
  doi={10.1109/TAC.2005.846556}
  }

@book{strikwerda2004finite,
author = {Strikwerda, John C.},
title = {Finite difference schemes and partial differential equations, Second Edition},
publisher = {Society for Industrial and Applied Mathematics},
year = {2004},
doi = {10.1137/1.9780898717938},
address = {},
edition   = {},
}

@book{evans2022partial,
  title={Partial differential equations},
  author={Evans, Lawrence C},
  volume={19},
  year={2022},
  publisher={American mathematical society}
}

@article{vasca2021practical,
title = {Practical consensus in bounded confidence opinion dynamics},
journal = {Automatica},
volume = {129},
pages = {109683},
year = {2021},
issn = {0005-1098},
doi = {https://doi.org/10.1016/j.automatica.2021.109683},
author = {Francesco Vasca and Carmela Bernardo and Raffaele Iervolino},
}

@article{motsch2014heterophilious,
author = {Motsch, Sebastien and Tadmor, Eitan},
title = {Heterophilious Dynamics Enhances Consensus},
journal = {SIAM Review},
volume = {56},
number = {4},
pages = {577-621},
year = {2014},
doi = {10.1137/120901866},
}

@article{demarina2015controlling,
  title={Controlling Rigid Formations of Mobile Agents Under Inconsistent Measurements},
  author={Garcia de Marina, Hector and Cao, Ming and Jayawardhana, Bayu},
  journal={IEEE Transactions on Robotics},
  volume={31},
  number={1},
  pages={31--39},
  year={2015},
  doi={10.1109/TRO.2014.2373145}
}

@article{demarina2016distributed,
  title={Distributed Rotational and Translational Maneuvering of Rigid Formations and Their Applications},
  author={Garcia de Marina, Hector and Jayawardhana, Bayu and Cao, Ming},
  journal={IEEE Transactions on Robotics},
  volume={32},
  number={3},
  pages={684--697},
  year={2016},
  doi={10.1109/TRO.2016.2559511}
}

@inproceedings{sun2014formation,
  title={Formation Movements in Minimally Rigid Formation Control with Mismatched Mutual Distances},
  author={Sun, Zhiyong and Mou, Shaoshuai and Anderson, Brian D. O. and Morse, A. Stephen},
  booktitle={Proceedings of the 53rd IEEE Conference on Decision and Control},
  pages={6161--6166},
  year={2014},
  organization={IEEE}
}

@article{sinha2022consensus,
  title={Consensus of Networked Double Integrator Systems Under Sensor Bias},
  author={Sinha, Pallavi and Srikant, Sukumar and Sinhmar, Himani},
  journal={International Journal of Adaptive Control and Signal Processing},
  volume={37},
  number={1},
  pages={278--297},
  year={2023},
  doi={10.1002/acs.3524}
}

@article{hendrickx2013cutbalanced,
  author={Hendrickx, Julien M. and Tsitsiklis, John N.},
  title={Convergence of Type-Symmetric and Cut-Balanced Consensus Seeking Systems},
  journal={IEEE Transactions on Automatic Control},
  volume={58}, number={1}, pages={214--218}, year={2013},
  doi={10.1109/TAC.2012.2203214}
}

\end{document}